\pdfoutput=1

\documentclass[review]{elsarticle}

\usepackage{lineno,hyperref}
\usepackage[utf8]{inputenc}
\usepackage[english]{babel}
\usepackage{amsmath}
\usepackage{amsfonts}
\usepackage{amssymb}
\usepackage{graphicx}
\usepackage{caption}
\usepackage{subcaption} 
\usepackage[left=2cm,right=2cm,top=2cm,bottom=2cm]{geometry}
\usepackage{setspace} 
\usepackage{placeins} 
\usepackage{lineno} 
\usepackage{multirow} 

\journal{Journal of Hydrology}

\biboptions{authoryear}
\begin{document}

\begin{frontmatter}

\title{Streamflow forecasting using functional regression}

\author[mymainaddress]{Pierre Masselot\corref{mycorrespondingauthor}}
\cortext[mycorrespondingauthor]{Corresponding author}
\ead{pierre-lucas.masselot@ete.inrs.ca}

\author[mysecondaryaddress]{Sophie Dabo-Niang}
\author[mymainaddress]{Fateh Chebana}
\author[mymainaddress,mythirdaddress]{Taha B.M.J. Ouarda}

\address[mymainaddress]{Centre Eau-Terre-Environnement (ETE), Institut national de la recherche scientifique (INRS),
490 de la Couronne, Québec, QC, G1K 9A9, Canada}
\address[mysecondaryaddress]{Université Charles-de-Gaulle Lille 3, Domaine du pont de bois, Laboratoire EQUIPPE, maison de la recherche, BP 149, 59653 Villeneuve d'ascq cedex, France}
\address[mythirdaddress]{Institute center for water and environment (iWATER), Masdar institute of science and technology,
PO Box 54224, Abu Dhabi, United Arab Emirates}

\begin{abstract}
Streamflow, as a natural phenomenon, is continuous in time and so are the meteorological variables which influence its variability. In practice, it can be of interest to forecast the whole flow curve instead of points (daily or hourly). To this end, this paper introduces the functional linear models and adapts it to hydrological forecasting. More precisely, functional linear models are regression models based on curves instead of single values. They allow to consider the whole process instead of a limited number of time points or features. We apply these models to analyse the flow volume and the whole streamflow curve during a given period by using precipitations curves. The functional model is shown to lead to encouraging results. The potential of functional linear models to detect special features that would have been hard to see otherwise is pointed out. The functional model is also compared to the artificial neural network approach and the advantages and disadvantages of both models are discussed. Finally, future research directions involving the functional model in hydrology are presented.
\end{abstract}


\end{frontmatter}


\section{Introduction}
\label{sec.intro}


Streamflow forecasting is an important topic in hydrology. Being able to precisely forecast streamflows is crucial for the adequate management of water resources systems. Therefore, there is a wide body of literature concerning streamflow forecasting, developing and applying a wide range of forecasting methods. For such a purpose, two types of models can be identified \citep{Fortin97}: a) physical models which apply deterministic equations to a set of input variables (such as physiographic features or rainfall) to obtain the desired streamflow values, and b) statistical models which model streamflows in a probabilistic way and which take into account the uncertainty in observed data. The latter are often cheaper to perform. The focus of this paper is on statistical models.

A large number of statistical models have been proposed for streamflow forecasting. Two major classes of such models can be distinguished: time series and regression models. The former is mainly based on the modelling of the streamflow autocorrelation structure while the latter focuses on the correlation between a response (streamflow) and explanatory variables (or covariates) regardless of the time structure. Hence, the forecasts of such explanatory variables are used to forecast future streamflows. Typical explanatory variables are precipitations and temperatures \citep{Sene09}. Some models are typically used to long term forecast such as linear regression \citep[\emph{e.g.}][]{Vogel99}, principal component regression \citep[\emph{e.g.}][]{Garen92,Eldaw03}, partial least square regression \citep[\emph{e.g.}][]{Tootle07} and wavelet regression \citep[\emph{e.g.}][]{Kisi09}. However, these models only forecast the streamflow volume at large time scales (usually seasonal or annual). This does not provide information about the shape of the hydrograph within these long periods, lacking accuracy about the moment when flows will be peaking.  

\begin{figure}[h] 
	\centering
	\begin{subfigure}[b]{.45\linewidth}
		\includegraphics[width=7cm]{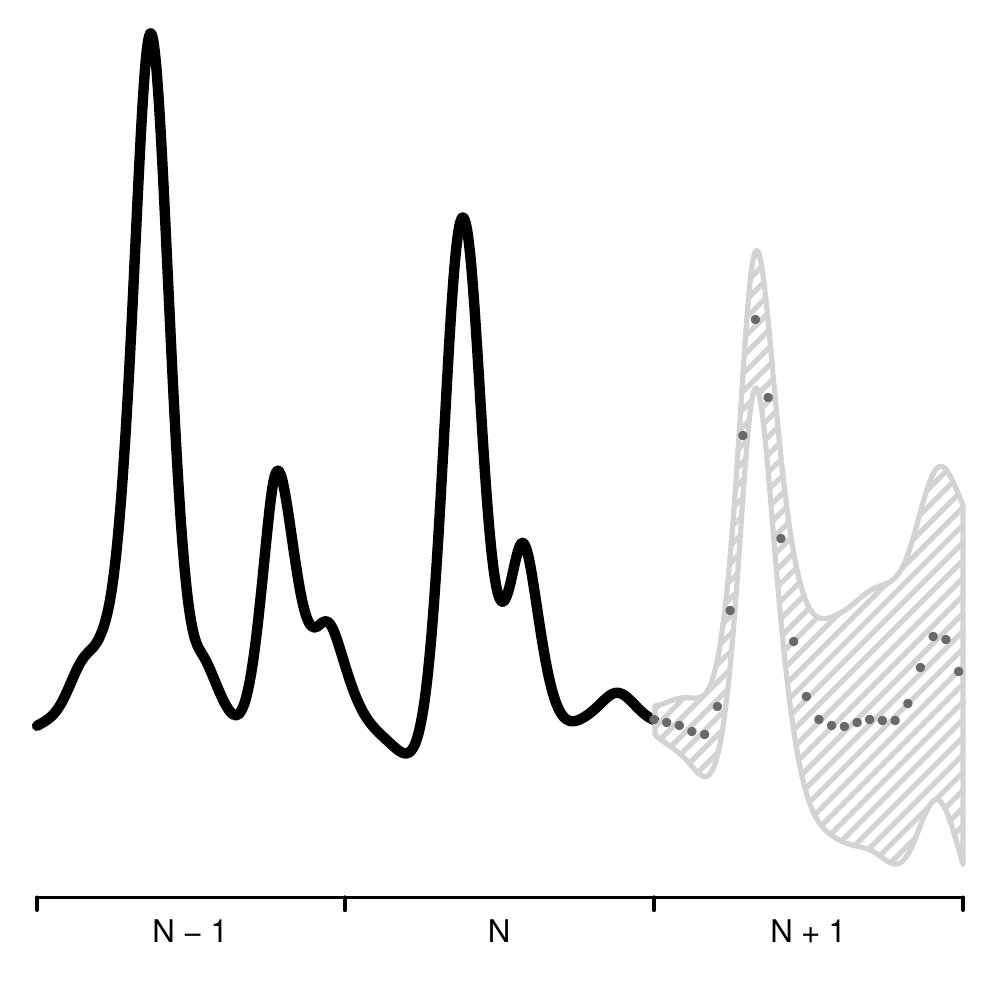} 
		\caption{Classical methods such as ANN or Wavelet regression.}
		\label{subfig.IllusNN}
	\end{subfigure}
	~
	\begin{subfigure}[b]{.45\linewidth}
		\includegraphics[width=7cm]{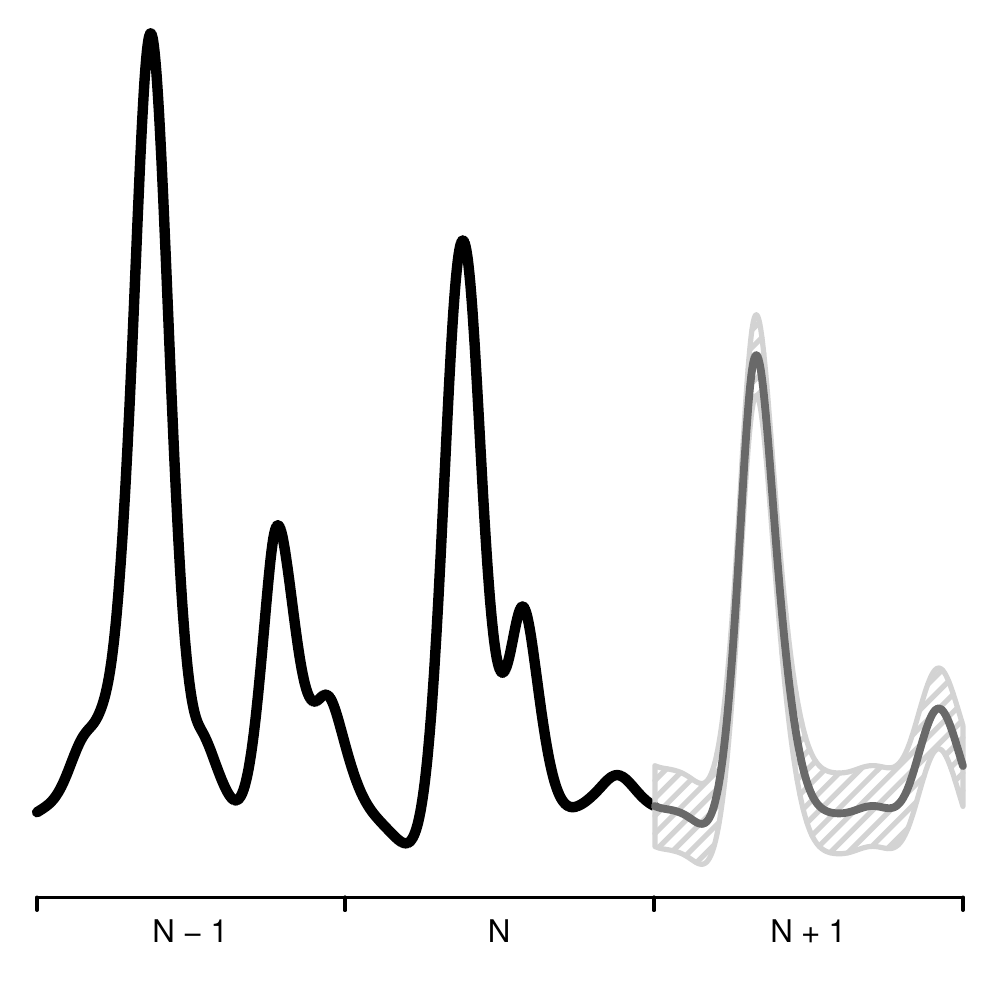} 
		\caption{Functional regression}
		\label{subfig.IllusFDA}
	\end{subfigure}
	
	\caption{Illustration of the difference between pointwise classical methods such as ANN models or wavelet regression and functional regression for forecasting purposes. The dashed area indicates the shape of expected confidence intervals for forecasts.}
	\label{fig.IllusForecast}
\end{figure}

There are also a number of models used to forecast short-term streamflows. The main models are artificial neural networks \citep[\emph{e.g.}][]{Zealand99,Kisi07,Chokmani08,Makkeasorn08}, wavelet regression \citep[\emph{e.g.}][]{Kisi09, Sahay13} and support vector regression \citep[\emph{e.g.}][]{Yu06}. However, these models can only model and predict few days (or few data points) at a time which is useful for detecting future extrema but not for forecasting overall trends in the streamflow process, such as the whole spring flood hydrograph for instance (such as illustrated in Figure \ref{fig.IllusForecast}).

In order to combine both short-term and long-term forecasting benefits, it is proposed to forecast streamflows using the whole streamflow process instead of single characteristics. This approach consists in considering the hydrograph as a continuous curve, \emph{i.e.} a \emph{functional data}, as proposed by \cite{Chebana12}. The statistical framework dealing with functional data is called \emph{functional data analysis} (FDA).  It was introduced by \cite{Ramsay82} and widely popularized by \cite{RamsayBook}. Such an approach became possible because a number of natural phenomena, such as streamflows, are now measured at a very fine scale or on a real time basis. This paradigm shift caused by FDA implies that the basic datum is now a curve instead of a scalar value. 

FDA has recently been introduced in hydrology by \cite{Chebana12} including the description and analysis framework of functional hydrographs and \cite{Ternynck15} used FDA to classify hydrographs. The present paper aims to introduce functional linear models (FLM) to hydrograph forecasting. A complete overview of FLM models is available in \cite{RamsayBook} as well as \cite{FerratyBook}. Among these models, two are particularly applicable for streamflow forecasting purposes: i) the scalar response FLM (FLM-S) expressing a traditional scalar variable according to functional variables and ii) the fully FLM (FLM-F) which aims at explaining a functional variable according to other functional variables. Table \ref{tab.flm} provides an overview of existing FLM models with references.

The FLM-S, which models a scalar response according to functional predictors, is perhaps the most studied model among FLM. It originated from \cite{Hastie93} and has known several theoretical developments since, \emph{e.g.}, \cite{Cardot99,Cardot03a}. The FLM-S is interesting because of its potential applications for long-term flood forecasting. This model is indeed well suited for modelling a hydrologic feature (such as a flood duration or volume) based on the complete history of the predictors (\emph{e.g.} temperatures or precipitations). This allows the use of more information than, for instance, classical linear regression. When using non functional predictors, a classical model with such fine information would be difficult to calibrate because it requires a number of lagged variables, resulting in a large number of coefficients to estimate, as well as a high collinearity between predictors. Thus, the uncertainty associated to such a classical model would be dramatically high. Since instead of simple scalar values the covariates of the FLM-S are curves, there is no need for many lagged variables and collinearity is no longer a problem \citep{Cuevas02}. 

The other model used in the present paper is the FLM-F introduced by \cite{Ramsay91} and studied theoretically by, \emph{e.g.}, \cite{Yao05}. An advantage of this model is its natural applicability to time series. Indeed, time series often contain features such as autocorrelation, trend or seasonality, creating spurious relationship and inducing wrong conclusions in classical linear models \citep[\emph{e.g.}][]{Granger74,Hoover03}. Even though several statistical methods have been developed to overcome this major drawback \citep[\emph{e.g.}][]{Phillips87}, linear models applied to time series always need a careful exploration of the considered time series. The FLM-F does not suffer from this drawback since a long sequence of data points can be considered as a single curve (such as a whole hydrograph). Moreover, its results are easier to interpret than many of the classical models used for day to day forecasting (\emph{e.g.}, ANN or support vector regression). The FLM-F model provides a way to forecast the overall trend of the hydrograph without a complex parametrization (illustrated by Figure \ref{fig.IllusForecast}). Finally, FLM-F allows changing the relationship between the predictor and the response over time, while classical models implicitly assume that the relationship does not change over time. Although it could be possible to fit a regression model without this assumption, this would result in a complicated model while it is a natural characteristic of FLM-F.

Globally, FDA is a topic receiving increasing attention in the theoretical and applied statistics literature. A number of reference textbooks dealing with functional data analysis already exist, such as \cite{Bosq00}, \cite{RamsayBook}, \cite{Bosq08}, \cite{DaboBook}, \cite{RamsayR} and \cite{Horvath12}. A large number of fields have already seen successful FDA applications, such as image processing \citep{Cardot03b}, medicine \citep{Ratcliffe02b,Ratcliffe02a}, genetics \citep{Muller08}, ecology \citep{Bel10}, marketing \citep{Sood09}, economy \citep{Goia10} and transportation systems \citep{Chiou12}. All these applications show the increasing interest in applying FLMs for practical purposes.

It is important to note that the expected improvements from using the functional framework may be mostly in terms of interpretation and statistical justification rather than raw performances. This includes more informations concerning the phenomena and conceptually more relevant results. We insist that functional regression methods are relevant since they are intuitively well suited and theoretically justified for time-related data. Indeed, these models represent a simple and clever way to exhibit many features of a phenomenon or a relationship between several variables.
 
The paper is organized as follows. After recalling the basics of data smoothing, functional linear models are introduced in section 2. Then, a case study on streamflow forecasting is presented in section 3. Section 4 concludes the study.
  
\section{Functional linear models}
\label{sec.background}

This section introduces FLMs starting with the necessary step of data smoothing. Then, the FLM-S and FLM-F are introduced and the procedure for their fitting by using empirical data is presented.  

\subsection{Data smoothing}
\label{subsec.smooth}

The main characteristic of a function $x(.)$ is its infinite dimensionality, meaning that there exists a value of $x(t)$ for each possible value of $t \in \mathbb{R}$ \citep{RamsayBook,FerratyBook}. Discrete measurements are thus not sufficient to represent such a datum, since they do not provide values between $x(t_j)$ and $x(t_{j+1})$. The usual way to represent an \emph{a priori} unknown function, is to express it as a sum of analytically known basis functions. This is actually the same thing as data smoothing, as it can be seen in other frameworks such as generalized additive models \citep{Hastie86,Chebana14}. However, in FDA it is a preliminary step aiming to prepare data to be functions while it is an aim itself in classical modelling. To obtain a set of $n$ functions $x_i(.), i=1,...,n$ with the same distribution (such as a collection of $n$ annual streamflow curves), the same basis functions must be employed for each curve. Hence, a functional datum $x_i(.)$ can be expressed as:
\begin{equation}
\label{eq.fd}
x_i(t) = \sum_{k=1}^{K_x} c_{ik} \phi_k(t) \; ; \; t \in \Omega 
\end{equation}
where $\phi_1(.),...,\phi_{K_x}(.)$ is a set of known basis functions. In practice the basis functions used are either Fourier basis when data are periodic, or B-spline when data are not periodic. However, other basis functions are possible such as wavelets which perform well when there are more local features. The relevance of the basis expansion approach lies in the fact that the problem of estimating a function is reduced to the estimation of a set of scalar coefficients $c_{ik}$. These coefficients are estimated by minimizing the penalized sum of square criterion:
\begin{equation}
\label{eq.sse}
PSSE = \sum_{j=1}^T \left( z_j - x_i(t_j) \right)^2 + \lambda \int D^2 x_i(t) \,dt
\end{equation}
where $D^2 x(t)$ represents the second derivative of $x(t)$, $z_{ij}, j=1,...,T$ are the measurements used to fit the function $x_i(.)$ and the $t_j$s are the corresponding measurement times. For instance, $z_{ij}$ could be the measured streamflow of a river on day $j$ of the year $i$ and $T=365$, meaning that $x_i(t)$ seeks to represent the flow over the $i^{th}$ year. Since observed data are usually noisy, in practice, the penalty term $\lambda \int D^2 x(t) \,dt$ is added to the sum of square criterion (\ref{eq.sse}) in order to penalize the rough $x_i(t)$. The parameter $\lambda$ controls the severity of the penalization, which means that the larger $\lambda$ is, the lesser the resulting number of basis $K$ will be. $\lambda$ is estimated by minimizing a cross-validation (CV) score which is an estimation of the prediction error, \emph{i.e.} the error we make when trying to predict new data \citep{Stone74}. These approaches are extensively explained and discussed in chapters 3 to 5 of \cite{RamsayBook}.

Finally, note that the formulation of curves as in (\ref{eq.fd}) allows the user to adapt the smoothing to catch special features of the curve. For instance, when special features of the curves need to be correctly estimated (such as peaks in frequency analysis), the use of B-spline basis functions with more breakpoints at locations where these features happen more often can prove to be useful. At these locations, the curve can thus adapt more easily to the targeted features. 

\subsection{Functional linear models}
\label{subsec.flm}

Now that the definition and construction of a functional data has been presented, this section introduces functional linear models, \emph{i.e.} how to model either a scalar or a functional response using functional predictors.

\subsubsection{Functional linear models for scalar response}
\label{subsubsec.scalar}

The functional linear model for scalar response seeks to explore the influence of a set of curves $X^{(j)}(t)$, $\left(j=1,...,q\right)$, at each time $t \in \Omega$ on a scalar response $Y$. To give an example, $Y$ can be the flow volume during a given period and the FLM-S gives the evolution of the influence of variables such as precipitations on the final flow volume. For simplicity and clarity purposes, the model with a single covariate $X(t)$ is presented. Nevertheless, we indicate at the end of the section how to generalize the model with several covariates.

The functional linear model for scalar response is, in the case of a single covariate, given by:
\begin{equation}
\label{eq.flm}
y_i = \alpha + \int_{\Omega} \beta(t) x_i(t) \,dt + \epsilon_i \; \;\;\;\;\;\;\;\; i=1,\ldots,n
\end{equation}
where $\left(x_i(.), y_i \right)$ are observed data from $\left(X(.),Y \right)$, $\alpha$ is the intercept term, $\beta(.)$ is the coefficient function, $\epsilon_i$ is the error term for observation $i$ and $n$ is the number of observations. Here the traditional regression coefficient is replaced by the coefficient function $\beta(t)$ which gives the influence of $x_i(t)$ on $y_i$ at each time $t$. For instance, in the ninth chapter of \cite{Ramsay09}, the model (\ref{eq.flm}) is used to fit the logarithm of total annual precipitation ($y_i$) according to temperature curves ($x_i(.)$) over a year (meaning that $\Omega = [0;365]$). This model can also be seen as a generalization of any linear model for which the curve $x_i(.)$ represents an infinity of scalar covariates.  

Since the coefficient $\beta(.)$ is a function, it is infinite dimensional. The functional model in (\ref{eq.flm}) has thus an infinity of degrees of freedom, which means that there is an infinity of solutions. Therefore, in order to reduce the degrees of freedom to a finite value, the coefficient $\beta(.)$ has to be expanded using a set of basis functions $\theta_k(t),\; k = 1,\ldots,K_{\beta} $, in the same manner as equation (\ref{eq.fd}), \emph{i.e.} $\beta(t) = \sum_{k=1}^{K_{\beta}} b_k \theta_k(t)$. The fitting of the FLM-S is then reduced to estimate a finite number of coefficients $b_k$.

Let $\phi_k(t),\; k = 1,\ldots,K_x$ be the set of basis functions used to represent the functional covariate $X(.)$. An observed curve from $X(.)$ is thus expressed as $x_i(t) = \sum_{k=1}^{K_x} c_{ik} \phi_k(t)$ similarly to equation (\ref{eq.fd}). Note that, since the curves $x_i(.)$ are observed, the coefficients $c_{ik}$ are estimated before fitting the FLM-S.

To explain how the coefficients $b_k$ are estimated, it is convenient to use the matrix notation of the basis expansion (\ref{eq.fd}) for both $\beta(.)$ and the set of observed curves $\mathbf{x}(.)=\left(x_1(.), \ldots , x_n(.) \right)$. They are thus expressed as: 
 
\begin{equation}
\label{eq.mat_expan}
\beta(t) = \Theta'(t)B \;\;\;\; \text{ and } \;\;\;\; \mathbf{x}(t) = C \Phi(t)
\end{equation}
where $\Theta(t) = \left( \theta_1(t),...,\theta_{K_{\beta}}(t) \right)'$ and $\Phi(t)= \left( \phi_1(t),...,\phi_{K_{x}}(t) \right)'$ are the vectors containing the basis functions evaluated at time $t$. $B$ is the vector containing the $K_{\beta}$ coefficients $b_k$ to estimate, and since $\mathbf{x}(t)$ is a set of $n$ curves, $C$ is a $n \times K_x$ matrix containing the coefficient $c_{ik}$ at line $i$ and column $k$.

The matrix form in (\ref{eq.mat_expan}) is convenient because it allows expressing the functional linear model in (\ref{eq.flm}), in the matrix form:
\begin{equation}
\label{eq.flm_matrix}
\mathbf{y} = \alpha + \int_{\Omega} C \Phi(t) \Theta'(t)B \,dt + \mathbf{\epsilon}
\end{equation}
for the vector $\mathbf{y}$ containing all observed scalar responses. Since $C$ and $B$ do not depend on $t$, it is possible to evaluate the $K_x \times K_{\beta}$ matrix $J_{\Phi \Theta} = \int_{\Omega} \Phi(t) \Theta'(t)\,dt$, and thus rewrite (\ref{eq.flm_matrix}) as $ \mathbf{y} = \alpha + C J_{\Phi \Theta} B  + \mathbf{\epsilon} $. Finally, by setting $\chi = \begin{bmatrix}
\mathbf{1} & C J_{\Phi \Theta}
\end{bmatrix}$ and $\mathbf{B}=(\alpha, B')'$, the functional linear model (\ref{eq.flm}) is written as : 
\begin{equation}
\label{eq.flm_final}
\mathbf{y} = \chi \mathbf{B} + \mathbf{\epsilon}
\end{equation}
which is similar to a traditional linear model with design matrix $\chi$ and coefficient vector $\mathbf{B}$. This model is fit by ordinary least squares.

The shape of the functional coefficient $\beta(.)$, and therefore the accuracy of the response forecasts depends on the value $K_{\beta}$ which controls the smoothness of the estimation. Thus, as for the smoothing of data curves, the smoothness of $\beta(.)$ can be controlled by directly choosing a small $K_{\beta}$ or by adding a roughness penalty to the SSE criterion. The latter allows to define a basis with a large number of functions, for which several coefficients are shrunk to zero. The criterion to be minimized is thus the penalized SSE:
\begin{eqnarray} \label{eq.pensse}
PENSSE_{\lambda}(\alpha,\beta) & = & \sum_{i=1}^n \left[ y_i -  \alpha - \int_{\Omega} \beta(t) x_i(t) \,dt  \right]^2 + \lambda \int _{\Omega} \left[L\beta(t) \right]^2 \,dt \nonumber\\
 & = & \| \mathbf{y} -  \chi \mathbf{B}  \|^2 + \lambda \mathbf{B}' R \mathbf{B}
\end{eqnarray}
where $L$ is a linear differential operator applied to $\beta(t)$ and 
$$R = \begin{bmatrix}
1 & \ldots \\
\vdots & \int_{\Omega} \left[L \Theta (t) \right] \left[L \Theta (t) \right]' \,dt
\end{bmatrix}$$ 
is a $(K_{\beta}+1) \times (K_{\beta}+1)$ penalization matrix. The first row and column of 1's are here to take into account the intercept $\alpha$. The model is now similar to a Ridge regression and thus has the same solution \citep{Hoerl70}. Traditionally, $L$ is defined as the second derivative (or acceleration) of $\beta(.)$ in order to penalize rough functions, but other differential operators are also possible, such as the harmonic acceleration \citep[][p. 93]{RamsayBook}. The coefficient $\lambda$ controls the severity of the penalization, and thus the smoothness of $\beta(.)$. In practice, $\lambda$ is chosen by minimizing the CV criterion \citep{Stone74}. 

The FLM-S is not tied to only one predictor and is generalized for several functional predictors and optional scalar predictors. There is also the possibility of using different time intervals for the covariates if the covariates have different time lags. In hydrology, this allows to consider variables such as snow height only during the months when they are not null (November to April in Canada). This generalization and more topics are covered in a number of reference textbooks \citep[\emph{e.g.}][]{RamsayBook,Horvath12}.

\subsubsection{Fully functional linear model}
\label{subsubsec.func_resp}

This section presents the fully functional linear model (FLM-F) which links the whole response curve to the whole covariate curve. This is a more general model than the ``concurrent'' model (also known as ``point-wise FLM'' or ``varying coefficient model'') developed for the case when response and covariate are both functional \citep[\emph{e.g.}][chapter 14]{Hastie93b,RamsayBook} but which links only time $t$ of the covariate to the same time $t$ of the response. Although an estimation method allowing the use of multiple functional covariates in the FLM-F is currently appearing \citep[see the working paper][]{Ivanescu13}, we present here the basic method with only one covariate of \cite{RamsayBook}, for simplicity purposes.    

The FLM-F is expressed as:

\begin{equation}
\label{eq.gen_flm}
y_i(t) =\alpha(t) + \int_{\Omega_1} \beta(s,t) x_i(s) \,ds + \epsilon_i(t) \; ; \; t \in \Omega_2
\end{equation}
where $\beta(s,t)$ is now a function of both $s$ and $t$, which means that it is a surface. Indeed, $\beta(s,t)$ gives the influence of $x_i(.)$ at time $s$ on $y_i(.)$ at time $t$, allowing also $\Omega_1$ and $\Omega_2$ to be different. This flexibility generalizes the use of lags in a linear model since it allows the significant lags to change over time. For instance, this model is able to take into account the fact that the lag between rainfall and the resulting streamflow can change over time.  

Note that the FLM-F includes the time domain $s>t$. When data are time series this means that there is an effect of the predictor on the response backwards in time. Attempts have been made at developing a ``historical FLM'' (HFLM) taking account only of the times $s<t$, \citep{Malfait03,Kim11}. However these models result in less smooth $\beta(.,.)$ surfaces.
Moreover, in practice the surface is very close to zero where $s>t$ (means no backward influence). Thus the present paper introduces the general case for clarity and simplicity purposes.

For the reason of infinite degrees of freedom exposed in section \ref{subsec.flm}, the surface $\beta(.,.)$ has to be expressed using basis functions. Since $\beta(.,.)$ is bi-dimensional, it is expanded using a tensor product expansion (also called outer product) which is a multi-dimensional version of (\ref{eq.fd}). The coefficient surface is thus expressed in terms of $K_1$ functions $\eta_k(s)$ and $K_2$ function $\theta_l(t)$, \emph{i.e.}:

\begin{equation}
\label{eq.double_basis_beta}
\beta(s,t) = \sum_{k=1}^{K_1} \sum_{l=1}^{K_2} b_{kl} \eta_k(s) \theta_l(t) = H(s)B \Theta(t)
\end{equation}
where $H(s)$ and $\Theta(t)$ are respectively the vectors containing the $K_1$ functions $\eta_k(s)$ and $K_2$ function $\theta_l(t)$. The coefficients to estimate are stored in the $K_1 \times K_2$ matrix $B$. 

As in the case of FLM-S, the FLM-F is fit through the minimization of a penalized sum of square criterion. Since the response is a function, the sum of squares is integrated, \emph{i.e.} the criterion to minimize is: 

\begin{equation}
\label{eq.ISSE}
ISSE(\alpha,\beta) = \int \sum_{i=1}^n \left[y_i(t)-\alpha(t) - \int_{\Omega_1} \beta(s,t) x_i(s) \,ds \right]^2 \,dt + PEN_0(\alpha) + PEN_1(\beta) + PEN_2(\beta)
\end{equation}
where $PEN_0(\alpha) = \lambda_0 \int_{\Omega_1}\left[L_0 \alpha(t) \right]^2 \,dt $ is a penalization on $\alpha$ while $PEN_1(\beta) = \lambda_1 \int_{\Omega_1} \int_{\Omega_2} \left[L_1 \beta(s,t) \right]^2 \,ds \,dt$ and $PEN_2(\beta)=\lambda_2 \int_{\Omega_1} \int_{\Omega_2} \left[L_2 \beta(s,t) \right]^2 \,ds \,dt$ are both penalizations on $\beta$. $L_0$, $L_1$ and $L_2$ are differential linear operators. Two distinct penalisation terms are used for $\beta(.,.)$ because of the tensor product which makes independent the two dimensions of the surface. This allows the complexity of the surface to be controlled independently on each dimension. As in the case of data smoothing and FLM-S, the parameters $\lambda_0$, $\lambda_1$ and $\lambda_2$ are chosen by minimizing a cross-validation criterion.

The fully functional linear model in (\ref{eq.gen_flm}) is fitted using the same heuristic as the FLM-S of section \ref{subsec.flm}. The ISSE criterion (\ref{eq.ISSE}) can be expressed in a matrix form and then be derived to obtain an equation to resolve (called the normal equation). Here, the quantity to estimate is a matrix instead of a vector in the FLM-S. The normal equations have to be rearranged using the Kronecker product in order to estimate the vector $vec(B)$ which is the matrix $B$ rearranged column-wise. The computational details are not shown here. They can be found in chapter 16 of \cite{RamsayBook}. 

An alternative way to provide an estimate for $\beta(.,.)$ in (\ref{eq.gen_flm}) was developed by \cite{Yao05}. This estimate uses the functional principal components \citep[FPC,][]{Ramsay82} as basis functions. More precisely, the FPCs of $X(s)$ are used to expand $\beta(s,t)$ on the $s$ dimension and the FPCs of $Y(t)$ on the $t$ dimension. The associated coefficients are proportional to the covariance between the FPC scores of $X(s)$ and $Y(t)$. Details of the procedure can be found in chapter 8 of \cite{Horvath12}. Although this latter estimation procedure has good theoretical convergence properties, it does not allow to control the smoothness of the estimation better than the procedure explained above. Moreover, the choice of the FPCs to consider is difficult \citep[\emph{e.g.}][]{Goldsmith11}. This is why the application presented here favors the approach of \cite{RamsayBook}.

Finally, as for the FLM-S, one can wonder if there are identifiability constraints on the number of basis for $\beta(s,t)$. In the present case, the inverted matrix is a Kronecker product of two matrices. This means that this matrix can be inverted if and only if the two matrices constituting the Kronecker product are not singular. It can be shown that this is the case, similarly to section \ref{subsubsec.scalar}, only if $K_1$ is lower than both $n$ and $K_2$.

\section{Application to streamflow}
\label{sec.appli}

Following the work of \cite{Chebana12} which described how to explore functional hydrographs, this section applies FLMs introduced in section \ref{sec.background} to forecast streamflows according to precipitation curves. In this application we focus on summer/autumn floods (floods caused by liquid precipitations) \emph{i.e.} streamflow data from July to October are considered for each year. To illustrate the FLM-S of section \ref{subsubsec.scalar}, the flow volume during the four mentioned months is modelled with the curves of precipitation as covariates. Then, the FLM-F of section \ref{subsubsec.func_resp} is used to predict the whole streamflow curve between July and October using the precipitation curve. Precipitations are considered from June to October in order to have past values for the first days of the streamflows time span. However, the model is applicable to the whole year. In fact, many possibilities exist with the functional linear model (e.g. only spring floods, only a month, etc…). 

\subsection{Data description}
\label{subsec.datadesc}
\begin{figure}[h] 
	\centering
	\includegraphics[width=15cm]{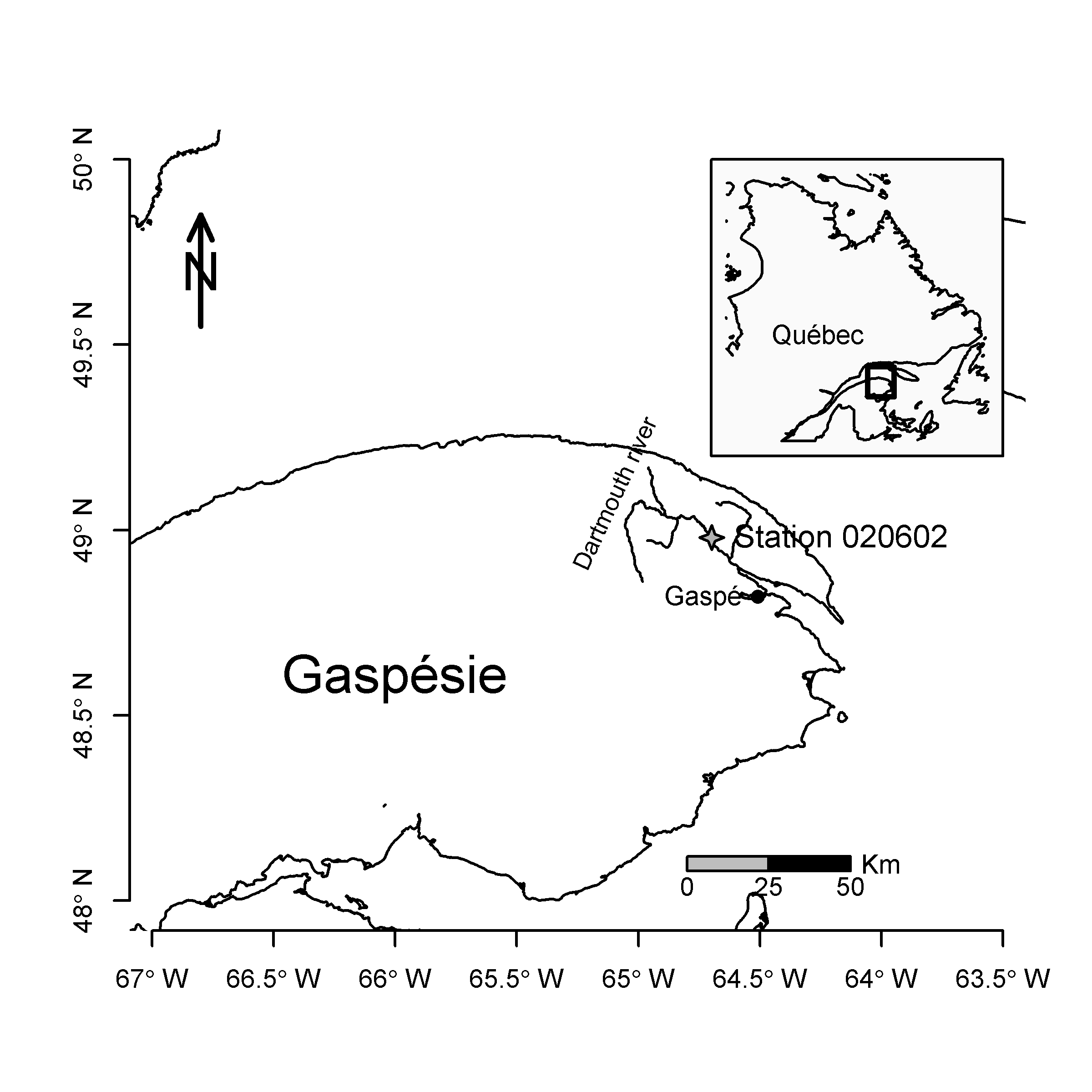} 
	\caption{Geographical location of the Dartmouth station.}
	\label{fig.map}
\end{figure}

The present application considers the Dartmouth station with federal reference number 01BH005. It is located on the Dartmouth river, 1.6 km upstream from the Ruisseau du Pas de Dame in the Gaspésie region of the province of Quebec, Canada. The drainage basin area is 626 $km^2$ and the flow regime of the river is natural. The geographical location of the station is shown in Figure \ref{fig.map}. The data consists in a daily streamflow series ($m^3 / s$) from 1970 to 2012 as well as daily total precipitation series available from 1981 to 2012. The common years between the series are thus 1981 to 2012. This means that, in the following, the sample size is $n=32$. The streamflow curves domain is the continuous interval $\Omega=\left[ 30;153 \right]$ and the precipitation time domain is $\Omega_p=\left[ 0;153 \right]$.

\cite{Chebana12} smoothed daily streamflows using 53 Fourier basis functions to obtain annual streamflow curves, which corresponds to a basis per week. Since the present application considers only the period July-October, the curves do not cover an entire year and Fourier basis appears less suited to smooth functional data. All the data used in this application (including precipitation data) are thus smoothed using B-spline basis which are more suited than Fourier basis for non periodic data.

Note that streamflow and precipitation data cannot physically have negative values, which means that there is an extra constraint in the smoothing of this data. Although it is possible to smooth under constraint \citep[see][chapter 6]{RamsayBook}, we choose here to smooth the logarithm of both streamflow and precipitation series. This has also the advantage of adding some symmetry around the mean curve for these series. In order to apply the logarithm for days with no precipitations, the values are set to 0.05 mm since the minimum value recorded is 0.2 mm. Note that the results are robust to the value attributed to the days without precipitation. Streamflow and precipitation curves are displayed in their original scale, \emph{i.e.} the inverse transformation is applied.

\begin{figure}[h] 
	\centering
	\begin{subfigure}[b]{.45\linewidth}
		\includegraphics[width=7cm]{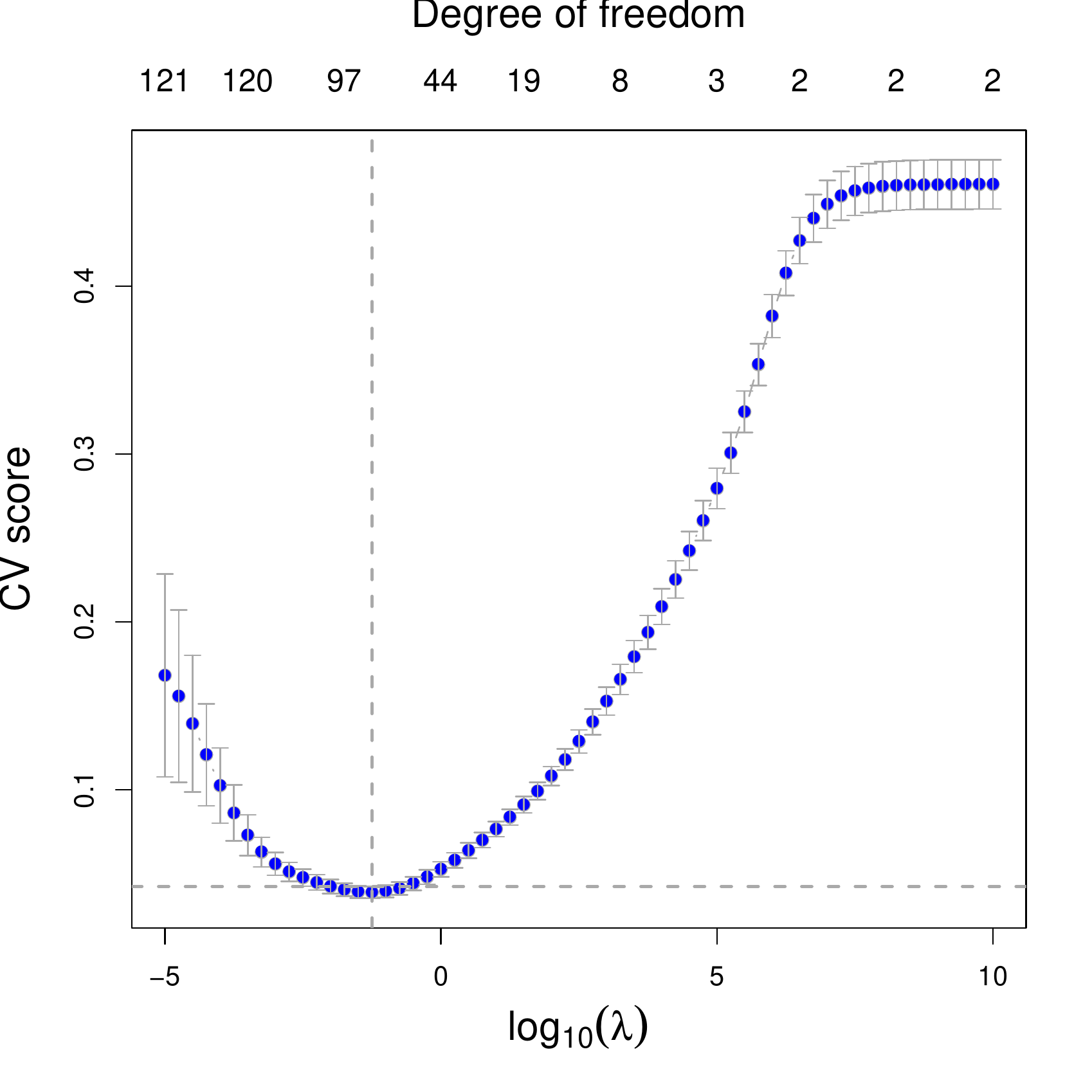} 
		\caption{Streamflows}
		\label{subfig.CVflow}
	\end{subfigure}
	~
	\begin{subfigure}[b]{.45\linewidth}
		\includegraphics[width=7cm]{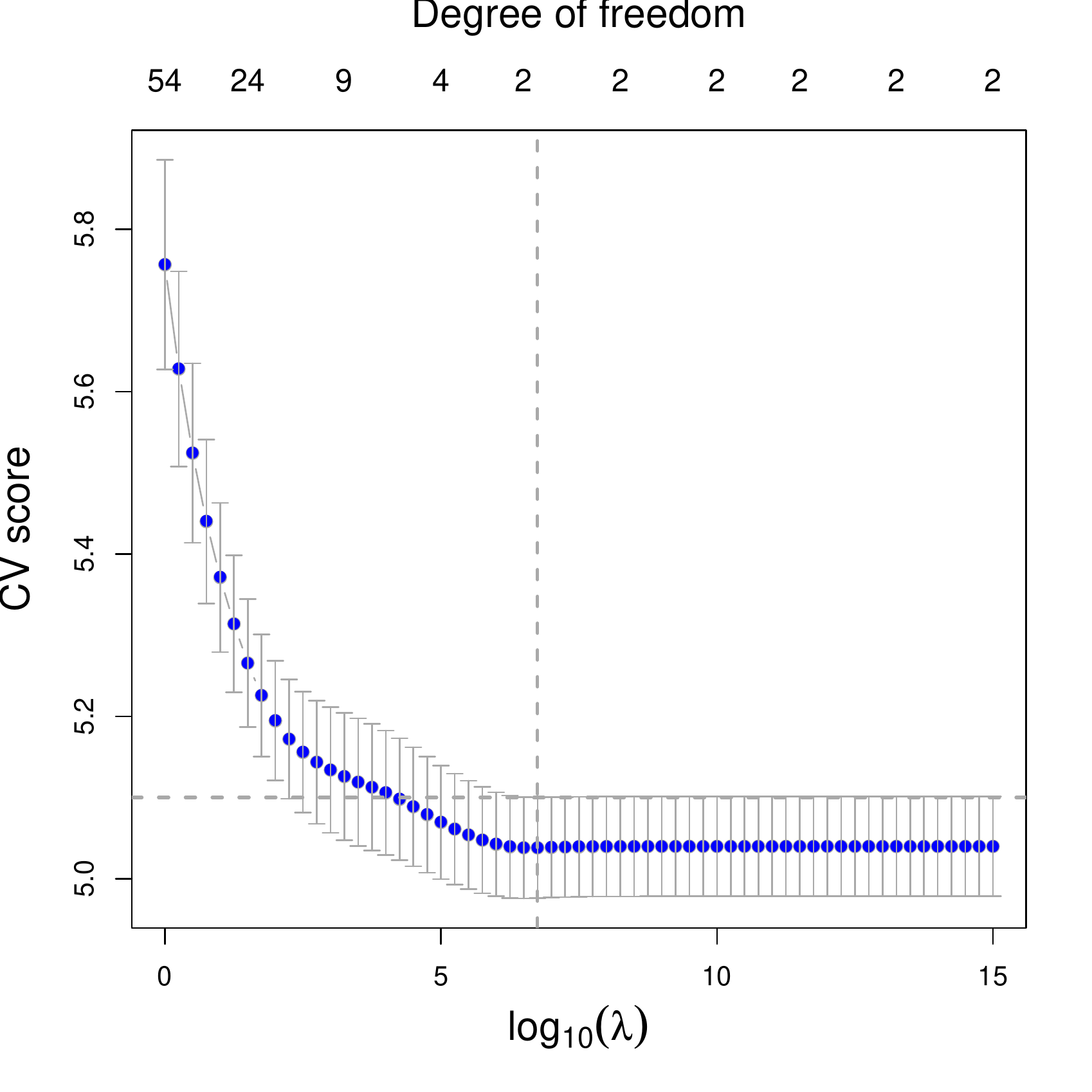} 
		\caption{Precipitations}
		\label{subfig.CVprec}
	\end{subfigure}
	
	\caption{10-fold CV curves for different values of $\lambda$ in the smoothing of streamflows and precipitation curves. The bars at each point represent the standard errors of the CV values. The vertical dashed line aims at spotting the minimum CV value and the horizontal dashed line is the minimum CV value plus its standard error. }
	\label{fig.cvSmooth}
\end{figure}

Figure \ref{fig.cvSmooth} shows the leave-one-out cross-validation (LOOCV) curve for choosing the parameter $\lambda$ in the smoothing of streamflows (for the FLM-F only) and precipitation curves (for both FLM-S and FLM-F). The smoothing of streamflow values is straightforward since a clear minimum appears for $\lambda = 10^{-1.25}$. This corresponds to 81 B-spline basis functions. The process is less straightforward for precipitations since the minimum is for a very high value of $\lambda$ which corresponds to only 2 basis functions. This translates the difficulty of predicting precipitation processes \citep{Suhaila11}. Since this would result in straight lines for precipitations, the $\lambda$ value chosen is the lowest with a CV value inside the standard error of the minimum (in other words, below the horizontal shaded line in Figure \ref{subfig.CVprec}). This results in $\lambda = 10^{4.25}$ and 7 basis functions. 

\begin{figure}[h] 
	\centering
	\begin{subfigure}[b]{.45\linewidth}
		\includegraphics[width=5cm]{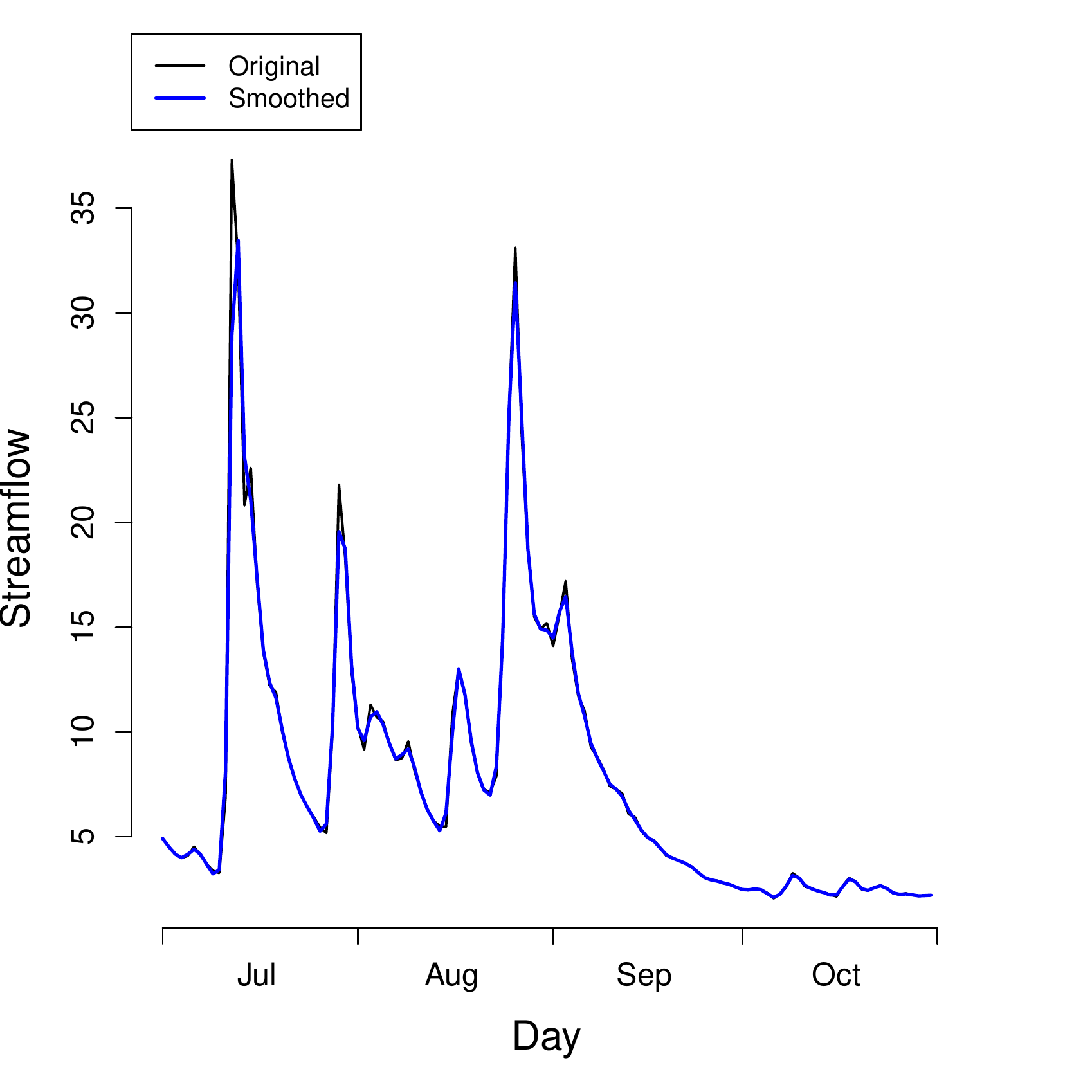} 
		\caption{Year 1989}
		\label{subfig.stream1989}
	\end{subfigure}
	~
	\begin{subfigure}[b]{.45\linewidth}
		\includegraphics[width=5cm]{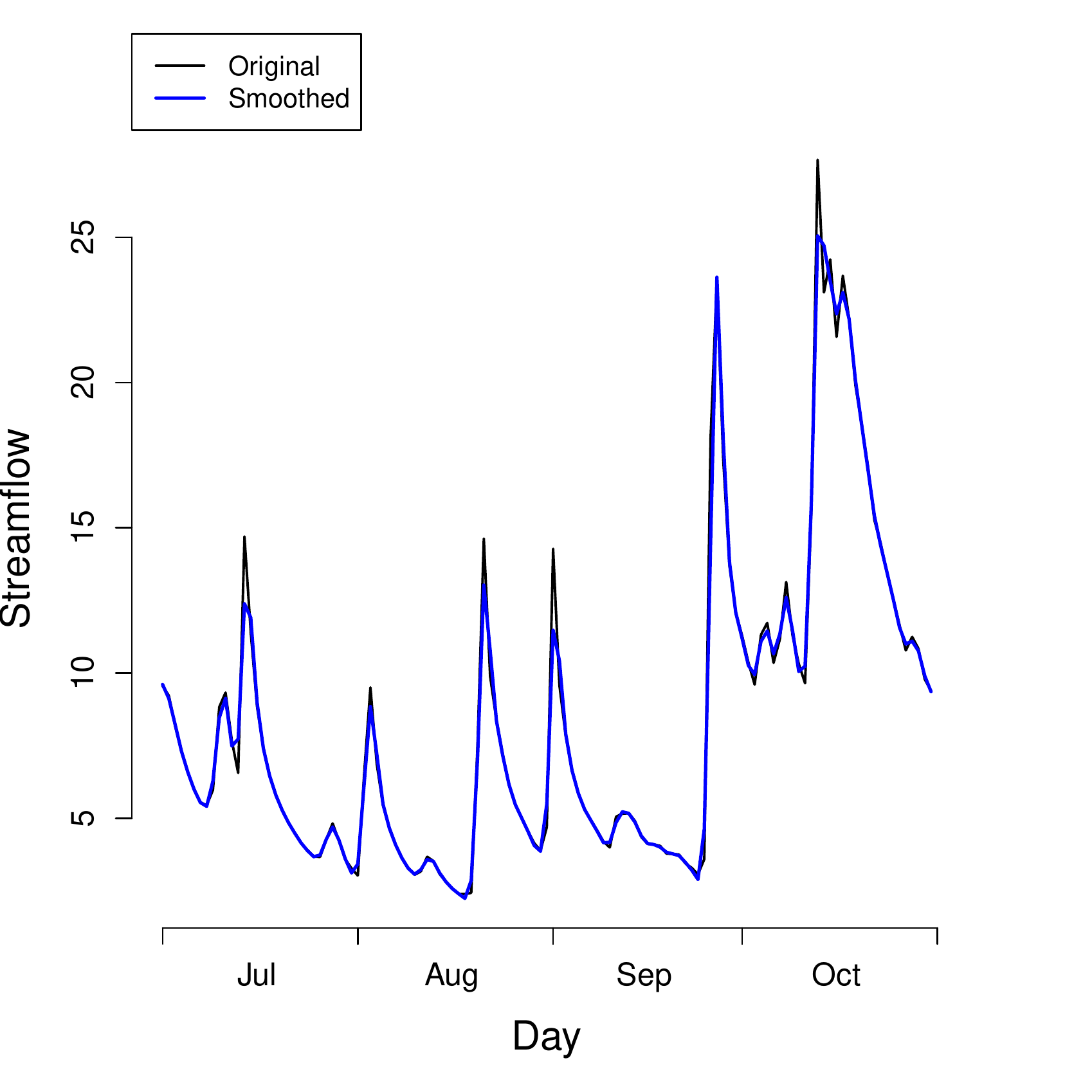} 
		\caption{Year 1991}
		\label{subfig.stream1991}
	\end{subfigure}
	~
	\begin{subfigure}[b]{.45\linewidth}
		\includegraphics[width=5cm]{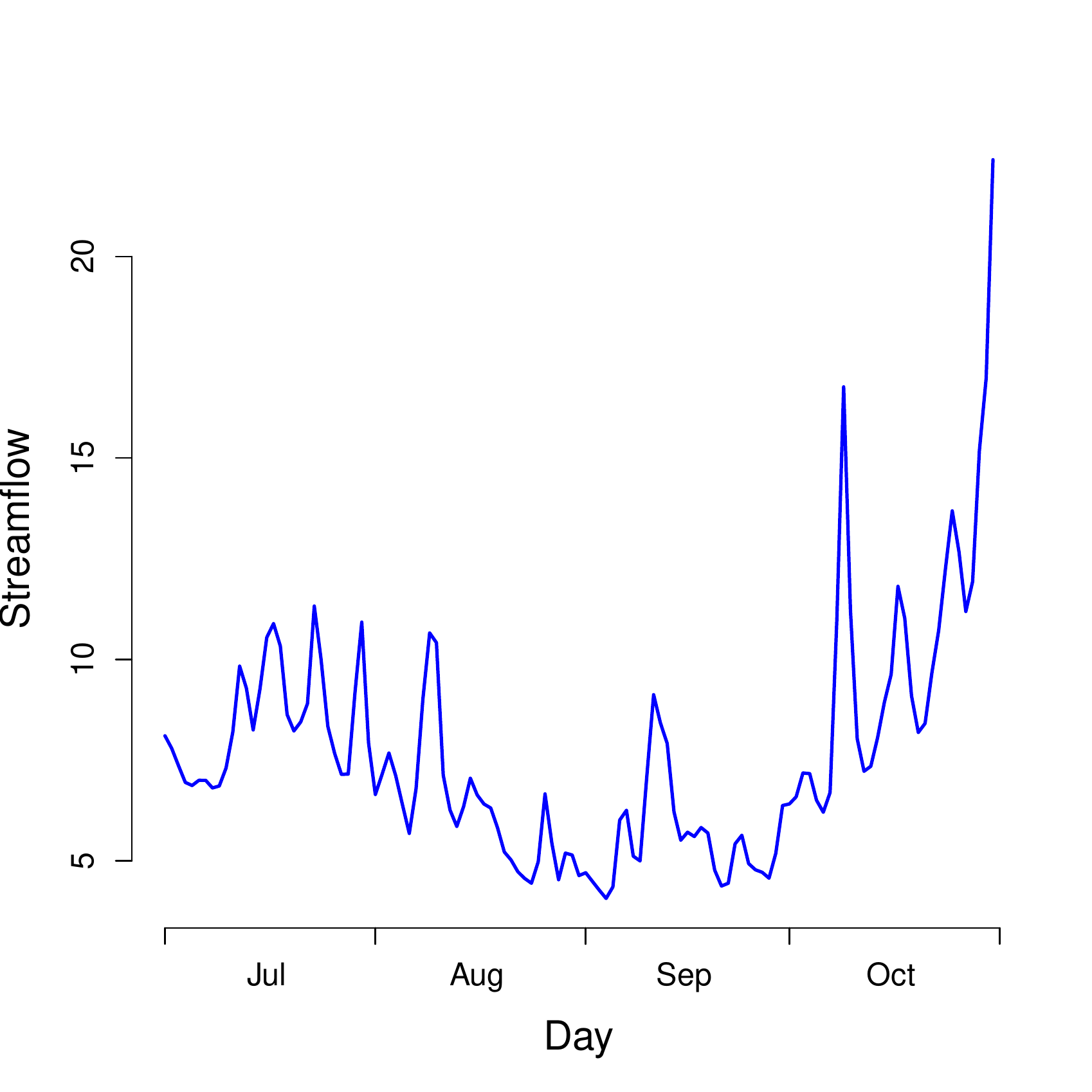} 
		\caption{Mean curve}
		\label{subfig.streammean}
	\end{subfigure}
	
	\caption{Examples of observed and smoothed streamflows for the years 1989, 1991 and the mean curve for the period 1981-2012. The black lines are observed streamflows and blue lines are the smoothed ones.}
	\label{fig.smoothed_flow}
\end{figure}

Figure \ref{fig.smoothed_flow} shows two examples of the resulting smoothed streamflows (with $\lambda = 10^{-1.25}$). Because of the high number of basis functions, the peaks are well reached by this smoothing. Peaks are probably the most important feature of streamflows and it is important that they are well represented. One look at the mean curve allows a characterisation of streamflows on this period. Indeed, it shows that streamflows are generally low during the middle of summer and begin to increase when the fall season begins. 
    
The same examples for precipitation curves are shown in Figure \ref{fig.prec}. It is the opposite of streamflows here since the curves are extremely smooth and only show periods when precipitations are more likely to happen. Indeed, a smoothing with more basis functions would have a prediction error that is too high. This means that precipitations are a phenomenon that is extremely difficult to predict. All the curves shown here are then used in the FLM applications of the next section.

\subsection{Forecasting of streamflow volume}
\label{subsec.volume}

For water resources management, it is important to know the amount of incoming water. Thus, it is of interest to estimate and forecast the total streamflow during a given period. This is often achieved through the use of regression models using the mean or total of some covariates \citep[\emph{e.g.}][]{Garen92,Eldaw03} or through regional frequency analysis \citep{Ouarda2000}. We illustrate in this section how to use the FLM-S for such a purpose.

Using the data described in section \ref{subsec.datadesc}, the method applied is the FLM-S (\ref{eq.flm}) described in section \ref{subsubsec.scalar}, with the variables
\begin{itemize}
	\item $y_i$: the response as the logarithm of the sum of daily streamflow values from the $1^{st}$ of July to the $31^{st}$ of October for year $i \; (i=1,...,n=31)$. The logarithm is used because total streamflows are strongly lognormal \citep{Vogel96};
	\item $x_i(.)$: the explanatory variable is represented by precipitation curves form the $1^{st}$ of June to the $31^{st}$ of October for year $i \; (i=1,...,31)$. 
\end{itemize}
Precipitations are considered up to one month earlier than streamflows in order to consider all precipitations that could influence the streamflow volume.

\begin{figure} 
	\centering
	\includegraphics[width=15cm]{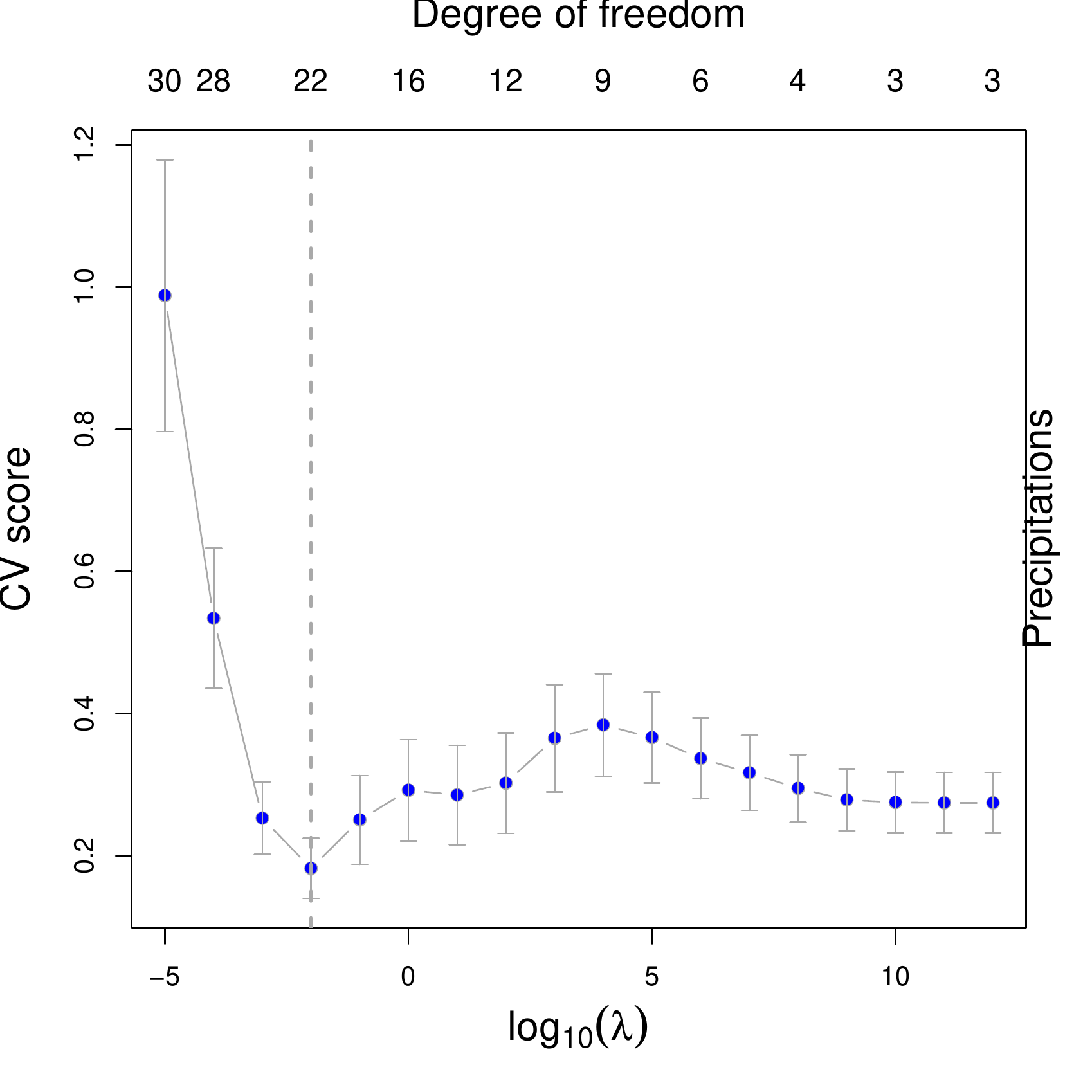} 
	\caption{Leave-one-out CV curve for the parameter $\lambda$ in the FLM-S. Bars indicate the standard error of the LOOCV values.}
	\label{fig.LOOCV_FLMS}
\end{figure}

To fit the FLM-S, the parameter $\lambda$ controlling the smoothness of the $\beta(.)$ coefficient curve must be chosen by minimizing the PENSSE criterion. Figure \ref{fig.LOOCV_FLMS} shows the LOOCV scores for different values of $\lambda$. There is an obvious minimum for $\lambda = 10^{-2}$, which corresponds to 22 basis functions on the B-spline basis and a rough $\hat{\beta}(.)$ curve.    

\begin{figure} 
	\centering
	\begin{subfigure}[b]{.45\linewidth}
		\includegraphics[width=5cm]{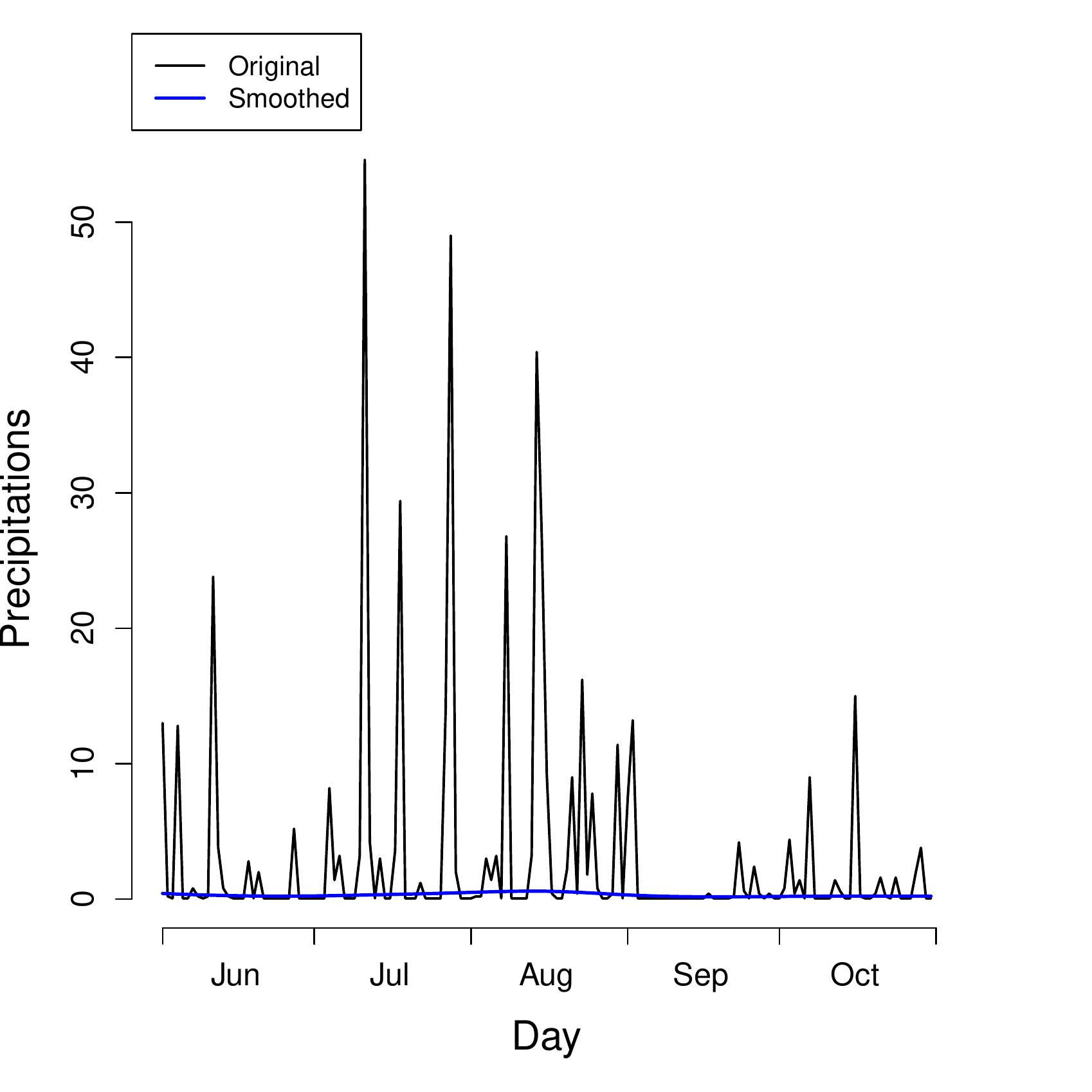} 
		\caption{Year 1989}
		\label{subfig.prec1989}
	\end{subfigure}
	~
	\begin{subfigure}[b]{.45\linewidth}
		\includegraphics[width=5cm]{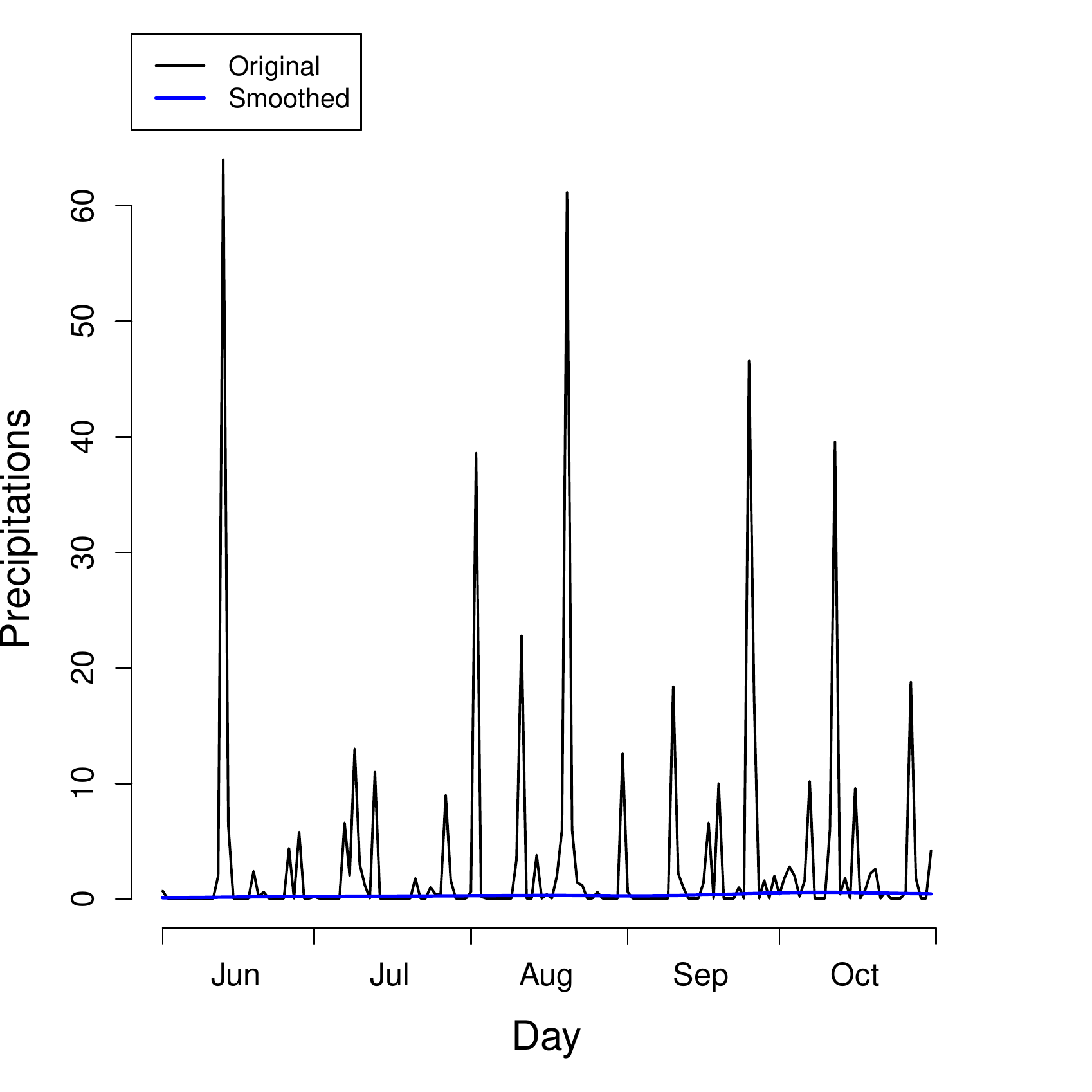} 
		\caption{Year 1991}
		\label{subfig.prec1991}
	\end{subfigure}
	~
	\begin{subfigure}[b]{.45\linewidth}
		\includegraphics[width=5cm]{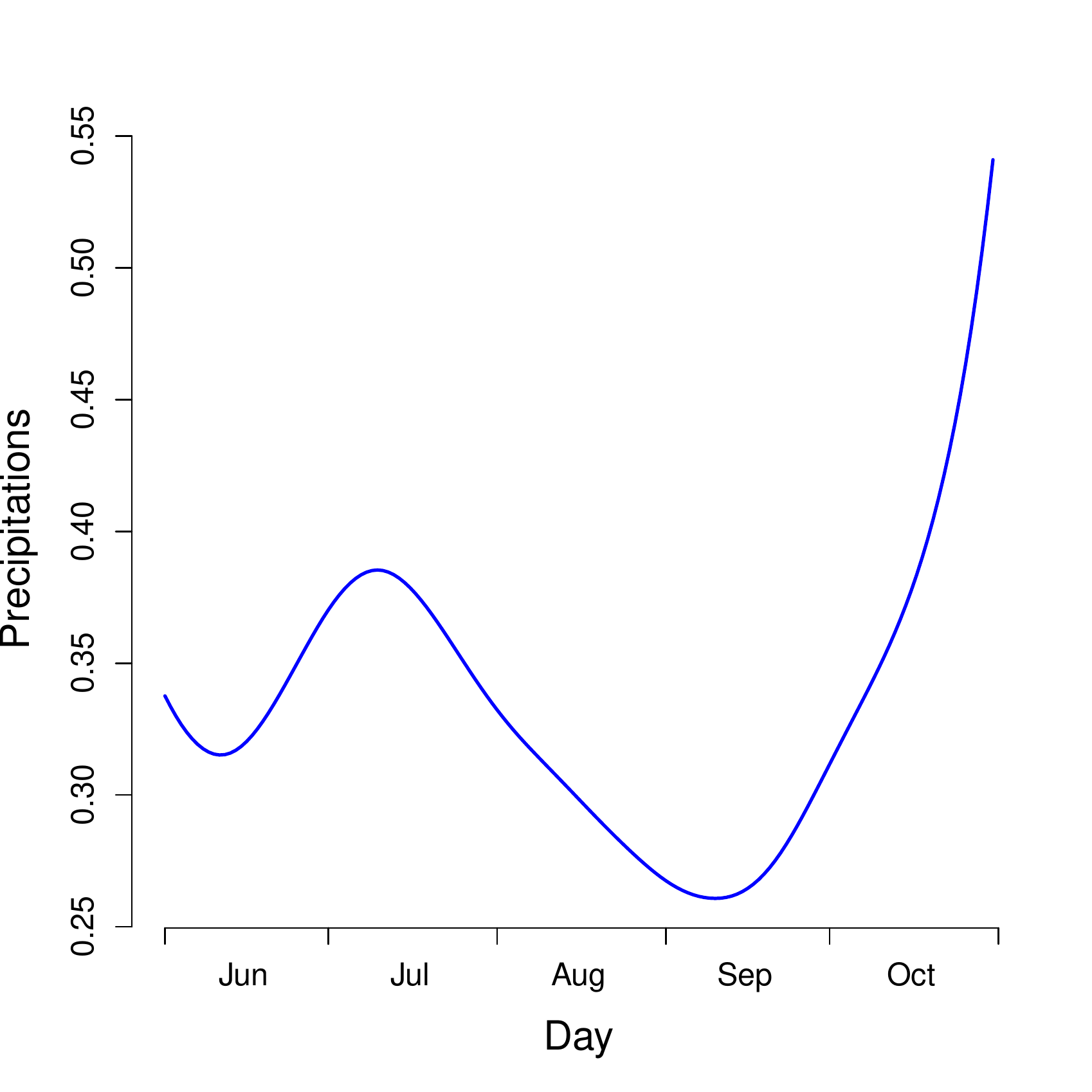} 
		\caption{Mean curve}
		\label{subfig.precmean}
	\end{subfigure}
	
	\caption{Examples of observed and smoothed precipitations for the years 1989, 1991 and the mean curve for the period 1981-2012. The black lines are observed precipitations and blue lines are the smoothed ones.}
	\label{fig.prec}
\end{figure}

\begin{figure} 
	\centering
	\includegraphics[width=15cm]{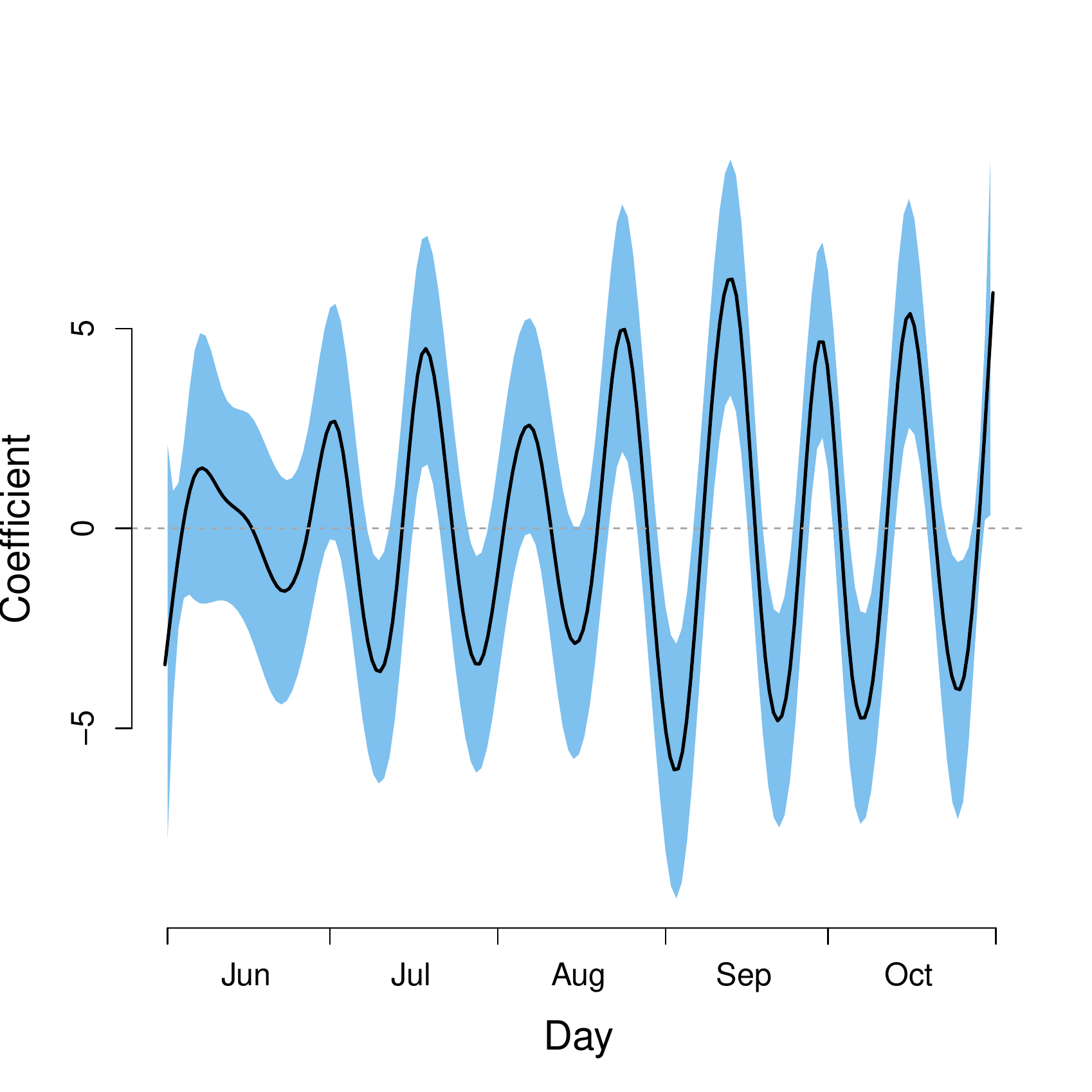} 
	\caption{Estimated $\hat{\beta}(.)$ function for precipitations in the FLM-S. The blue area corresponds to the pointwise 95\% confidence interval.}
	\label{fig.beta_FLMS}
\end{figure}

The $\hat{\beta}(.)$ curve obtained by fitting the FLM-S with $\lambda = 10^{-2}$ is shown in Figure \ref{fig.beta_FLMS}. The 95\% confidence interval is obtained by estimating a standard error curve and multiplying it by the quantiles of the standard gaussian distribution \citep[p.141]{RamsayR}. The low amplitude of the $\hat{\beta}(.)$ curve is due to the spreading of the influence of precipitation over the entire time span. Moreover, recall that the response of this model is the log volume. The $\hat{\beta}(.)$ curve shown in Figure \ref{fig.beta_FLMS} is not smooth and looks like an oscillation. This is because the data number ($n = 32$) is small and the $\hat{\beta}(.)$ curve is sensitive to the small features in precipitation data. However, it shows two periods where the oscillations have a larger amplitude : the second half of July and the beginning of the fall, indicating two periods where streamflows are more influenced by precipitations. At the end of July, streamflows can be low because the snow melt is over and the river has almost dried out. Therefore, any rainfall has a large influence on streamflows. Moreover, the large amplitude period towards the end of July also quickly follows the rainy period which occurs at the beginning of the summer, as indicated by Figure \ref{subfig.precmean}. The same reason can be evoked for the fall period of the $\hat{\beta}(.)$ curve which aligns with the beginning of autumn rainfalls (Figure \ref{subfig.precmean}).

In order to assess the performances of the FLM-S, the fit with $\hat{\beta}(.)$ is compared to a traditional linear model with the same response variable (denoted ``LM'' in the following). The explanatory variable of LM is the sum of daily precipitations between the the $1^{st}$ of June to the $31^{st}$ of October. This model results in a coefficient that is equal to $0.004$ which is significantly different from 0 ($p-value=5e^{-6}$). This coefficient is a kind of integration of the $\hat{\beta}(.)$ curve, the latter can be seen as the detail of the LM coefficient. 

\begin{figure} 
	\centering	
	\includegraphics[width=15cm]{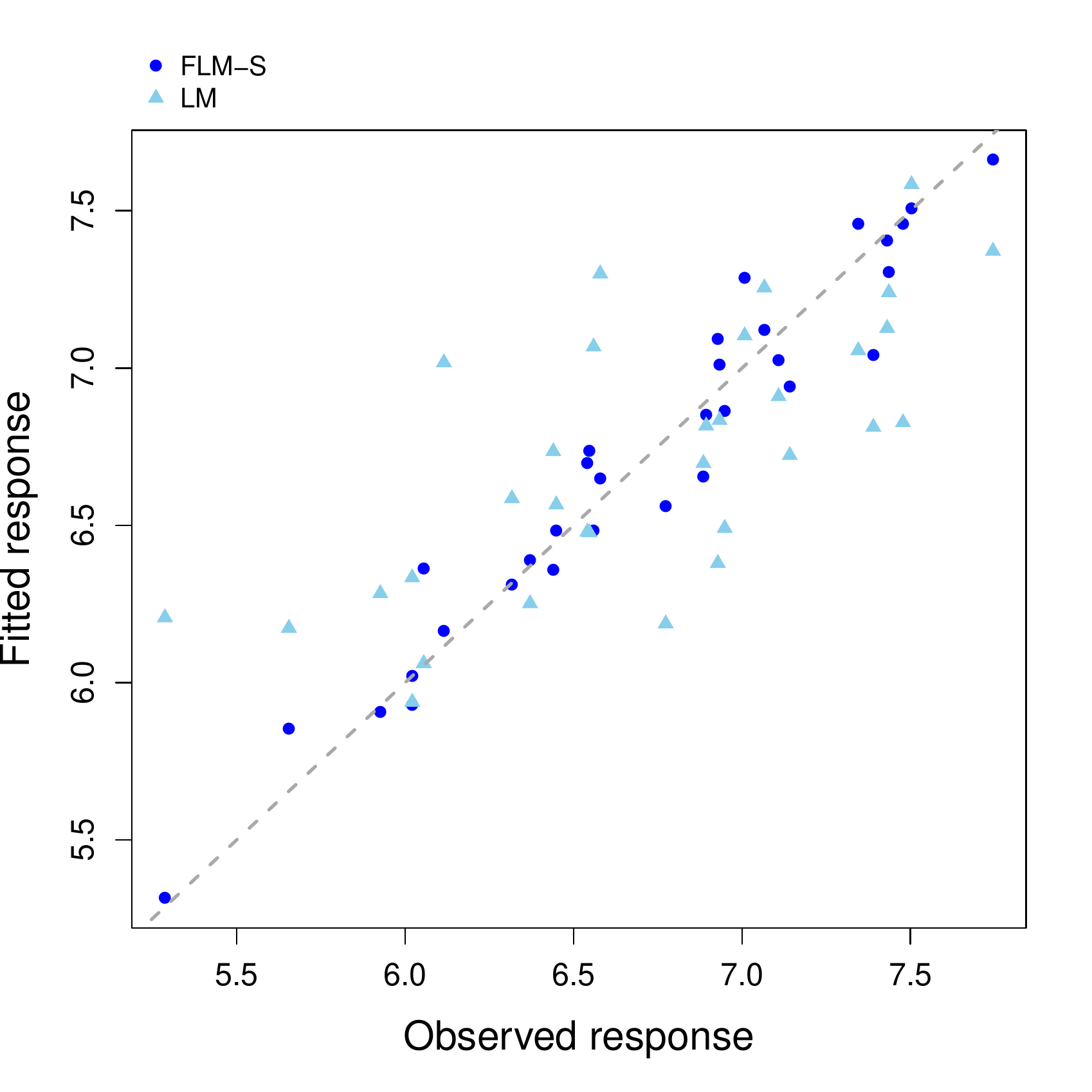} 
	\caption{Scatterplot of fitted vs. observed response values for the sum of streamflows. The plot contains the fitted values of the models FLMS (blue circles) and LM (cyan triangles).}
	\label{fig.obs_vs_fit}
\end{figure}

The scatterplot of Figure \ref{fig.obs_vs_fit} compares the fitted values $\hat{y}_i$ of the FLM-S and the LM. The fit is visually better for the FLM-S since the points are closer to the $y=x$ line (which represents a perfect fit). Table \ref{tab.criteria} illustrates the scores for different performance indicators (RMSE, R$^2$, LOOCV and bias). It also shows better performances for the FLM-S. Indeed, the FLM-S displays a higher R$^2$ and a lower RMSE than the LM. Note that the R$^2$ is close to one for the FLM-S (equal to 94\%), indicating an excellent fit. Note also that the difference between the two models is smaller with the LOOCV criterion. The complexity of the FLM-S seems somehow to balance its better performances. However, this should vanish with longer data records. Finally, note that the bias of the FLM-S is not null because of the regularization used for the fit. It is however close to zero and very small compared to the scale of the response (it accounts for less than 0.01\% of the mean of the response).


\subsection{Hydrograph forecasting}
\label{subsec.annualcurve}

\subsubsection{FLM-F fitting}
In this application, we are not interested in a single feature of the streamflow process but rather in the whole hydrograph. The objective is to forecast the hydrograph using the FLM-F given by (\ref{eq.gen_flm}). Although recent developments of the FLM-F allow its use with several covariates \citep{Ivanescu13}, only precipitations are used here for simplicity purposes. 

A proper estimation of the FLM-F needs the choice of three regularization parameters, \emph{i.e.} one for the intercept curve $\alpha(.)$ ($\lambda_0$), and two for the coefficient surface $\beta(.,.)$ ($\lambda_1$ for the $s$ dimension and $\lambda_2$ for the $t$ dimension). The resulting 3-dimensional plot is not shown here but the LOOCV (which, in this case, is a "leave-one-year-out" CV) is minimized when the three parameters are $\lambda_0=\lambda_1=\lambda_2=10^5$. 

\begin{figure} 
	\centering
	\begin{subfigure}[b]{.45\linewidth}
		\includegraphics[width=7cm]{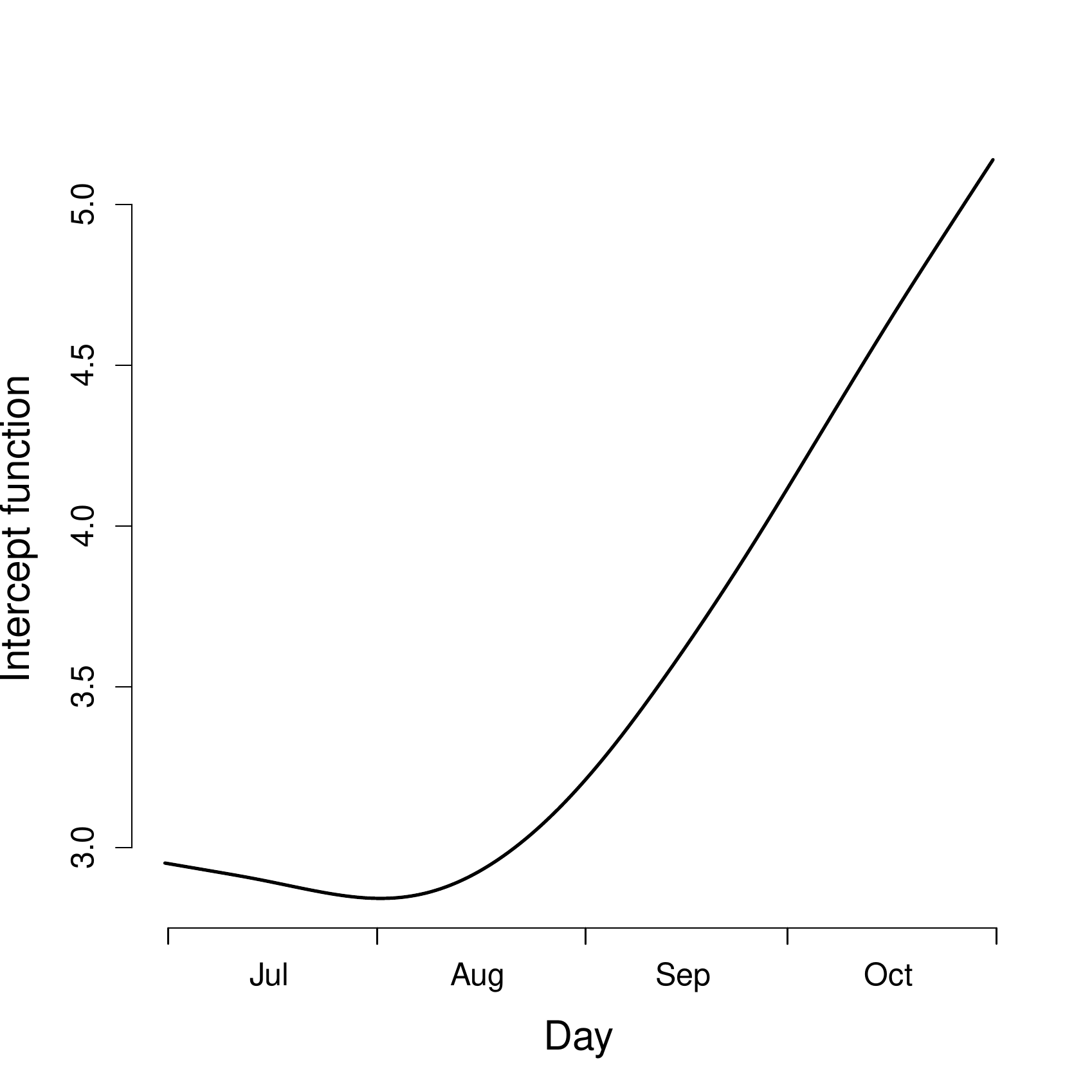}
		\caption{Intercept curve $\hat{\alpha}(.)$}
		\label{subfig.alphacurve}
	\end{subfigure}
	~
	\begin{subfigure}[b]{.45\linewidth}
		\includegraphics[width=7cm]{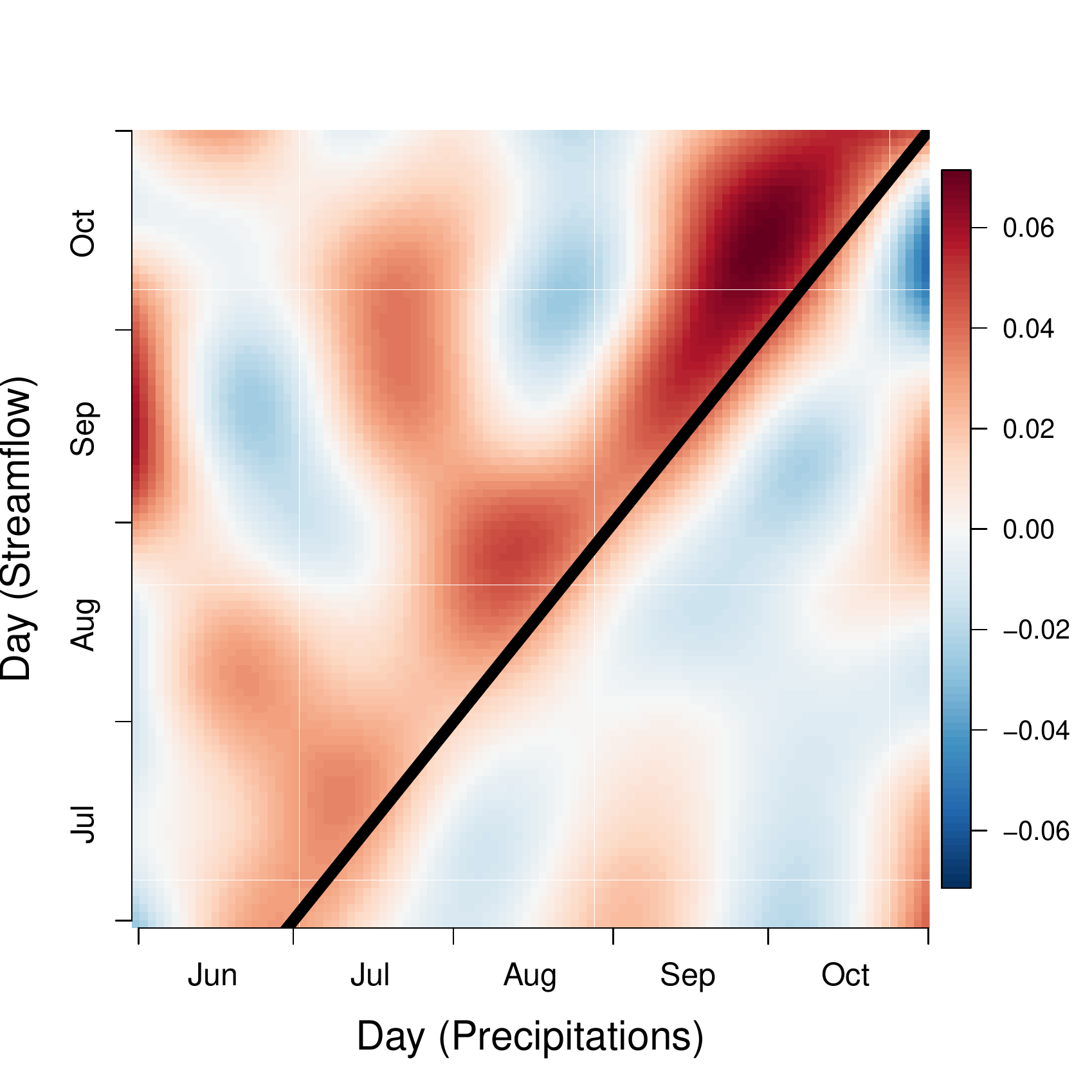}
		\caption{Coefficient surface $\hat{\beta}(.,.)$}
		\label{subfig.betasurf}
	\end{subfigure}
	\caption{Estimated functional coefficients of the FLM-F. The $\hat{\alpha}(.)$ curve shows the expected shape of the hydrograph without any influence of precipitations and the $\beta(s,t)$ surface shows the influence of precipitations on the hydrograph. For the latter, the dimension $s$ is in the abscissa and the dimension $t$ in the ordinate. The thick black line indicates the times $s=t$}
	\label{fig.betasurface_smooth}
\end{figure}

The estimated coefficients $\alpha(.)$ and $\beta(.,.)$ are shown in Figure \ref{fig.betasurface_smooth}. The $\alpha(.)$ curve illustrates the shape of the hydrograph without the influence of precipitations. The curve reaches its minimum during the middle of the summer (the beginning of august) and rises afterwards when the fall season starts. The $\beta(.,.)$ surface is a bump on the diagonal which indicates the positive influence of precipitation on the hydrograph, with a slight translation to the left to show delay between rainfall and the increase of streamflow. This influence is stronger during autumn season when more floods occur. This can be explained by the fact that there is less evaporation during this period of the year when temperature is not high. There is also a little extension of the bump to the top of the surface during July which could mean that, since the ground is dry at this time, infiltration is larger which leads to a significant contribution to surface flows two months later.

\subsubsection{Comparison with the artificial neural network approach}
To understand the strengths and weaknesses of the FLM-F, it has to be compared to other commonly used forecasting methods using exogeneous covariates in hydrology. Among these methods (reviewed in the introduction), ANN models are the most widely used because of their ability to simulate complex relationship. Thus, for comparison purposes, ANN models are considered to forecast the log streamflow as well. Following number of previous references \citep[\emph{e.g.}][chapters 1 and 2]{Anctil04,Chau05,Yonaba10,Govindaraju10}, ANN models are used with the 3 previous days of precipitations as covariates. For more generality, the covariate lag could have been selected using a cross validation procedure \citep[\emph{e.g.}][]{Haddad13}. However, for comparison purposes, the model design is based on what is found in the literature. The ANN estimation is made with one hidden layer containing 5 nodes \citep[the number of nodes is chosen using the leave-one-out cross-validation such as in][for instance]{Wu10}.

\begin{figure} 
	\centering
	\includegraphics[width=15cm]{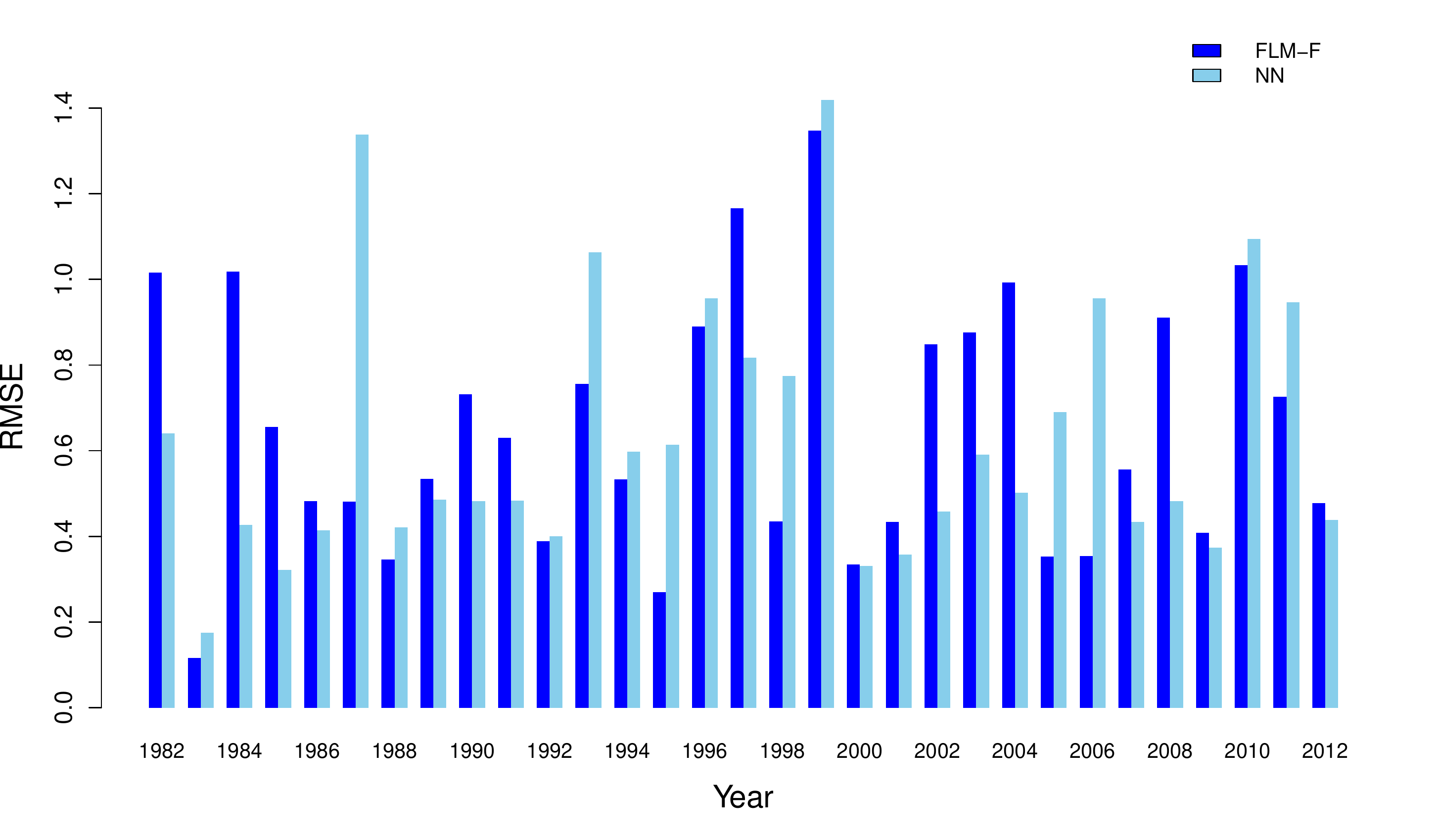} 
	\caption{Mean prediction error over all data points for each year.}
	\label{fig.FLMFCVyears}
\end{figure}

\begin{figure} 
	\centering
	\begin{subfigure}[b]{.45\linewidth}
		\includegraphics[width=7cm]{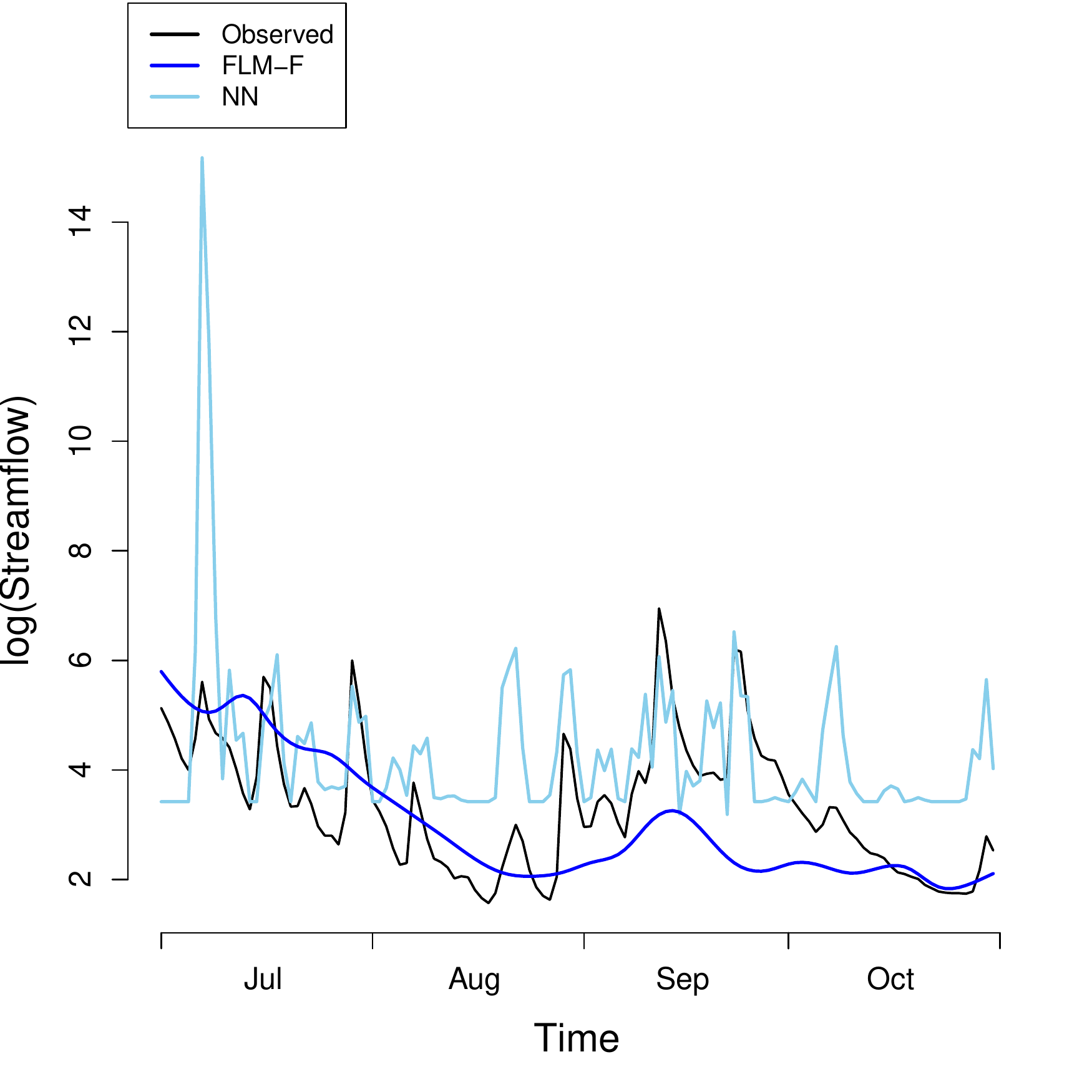} 
		\caption{Year 1983}
		\label{subfig.FLMFpred1983}
	\end{subfigure}
	~
	\begin{subfigure}[b]{.45\linewidth}
		\includegraphics[width=7cm]{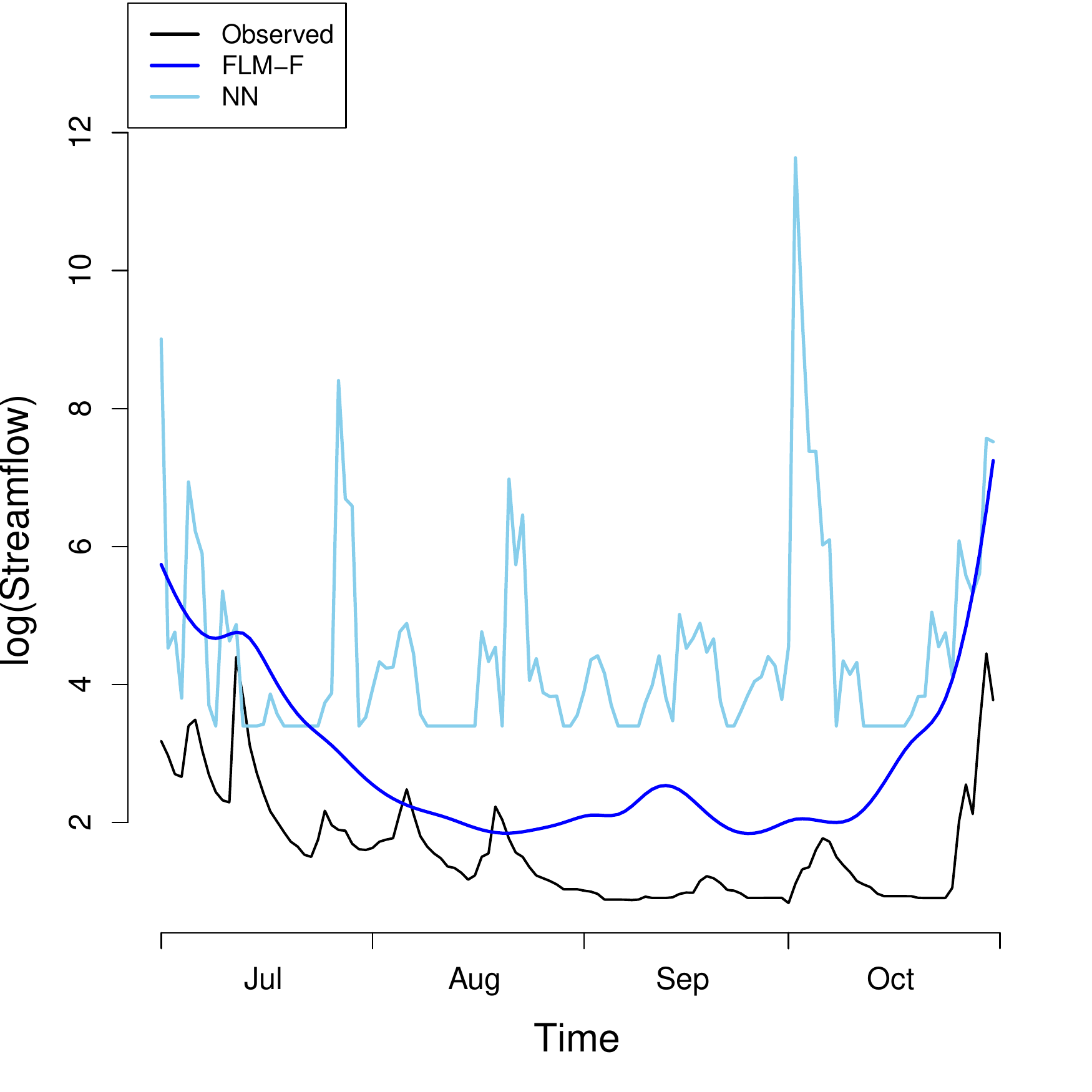} 
		\caption{Year 2009}
		\label{subfig.FLMFpred1987}
	\end{subfigure}
	~
	\begin{subfigure}[b]{.45\linewidth}
		\includegraphics[width=7cm]{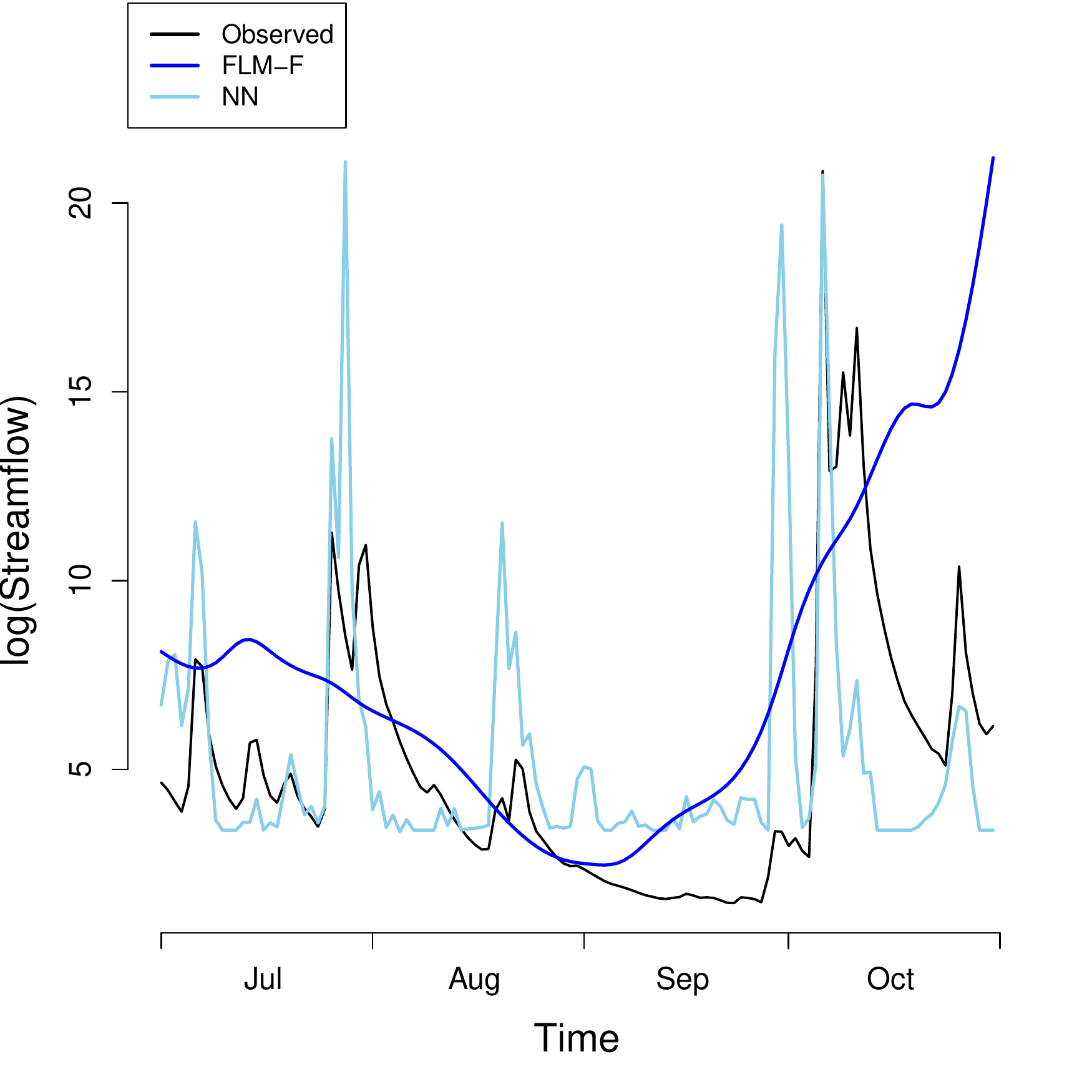} 
		\caption{Year 2009}
		\label{subfig.FLMFpred2009}
	\end{subfigure}
	\caption{Predicted hydrograph with the estimated FLMF and the estimated ANN model. For each year, the model is fitted on every other years and the remaining year is predicted.}
	\label{fig.FLMFpred}
\end{figure}

Figure \ref{fig.FLMFCVyears} shows the mean prediction error estimated by CV for each year. Performances of ANN models and FLM-F depend on the year, and it is hard to discriminate the two methods using only this figure. Three particular examples of predicted hydrographs $\hat{y}_i(.)$ using the FLM-F and the ANN models are shown in Figure \ref{fig.FLMFpred} to help understand when the FLM-F performs better than ANN models. It is immediately visible that the FLM-F and the ANN approaches have very different behaviours. The FLM-F actually predicts the global shape of the hydrograph, while the ANN model predicts only the short term patterns such as the peaks, but does not forecast well the streamflow accumulations. Hence, it appears clear that the FLM-F model is suited to match the trend and hence will perform better for years with few peaks. Moreover, FLM-F seems able to match some behaviours such as July droughts (\emph{e.g.} Figure \ref{subfig.FLMFpred1987}). However, when there are more short-term features such as peaks, the ANN performs better than the FLM-F (\emph{e.g} Figure \ref{subfig.FLMFpred2009}). The FLM-F forecast smoothness is mainly due to the smoothness of the $\beta(.,.)$ surface. The consideration of less smooth precipitation curves and $\beta(.,.)$ surface increases the prediction error since precipitations are extremely difficult to predict with precision. Note that the small number of years of record (32 years) does not allow for a rough $\beta(.,.)$ surface and the FLM-F should improve with more data years.

\begin{figure} 
	\centering
	\includegraphics[width=15cm]{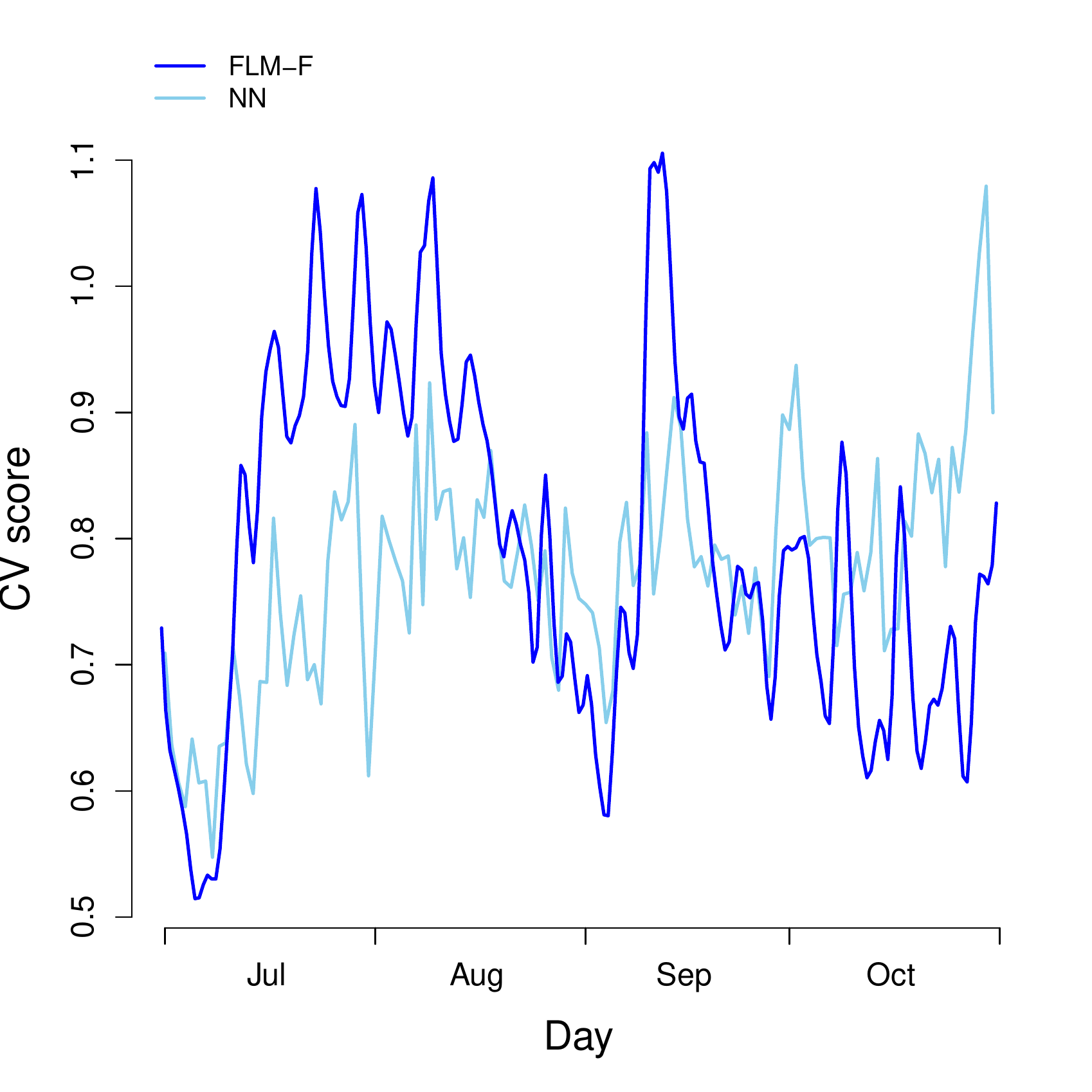} 
	\caption{Mean prediction error curve estimated by CV over all years. }
	\label{fig.FLMFCVcurvesMean}
\end{figure}

Table \ref{tab.criteria2} shows the global fitting criteria and indicates that the ANN model leads globally to better performances. Indeed, its RMSE value is lower and its R$^2$ value is higher than those of the FLM-F. However, the difference between the FLM-F and the ANN model is smaller for the CV score which is the only criterion that considers their performance on new data not used for the calibration of the models. This means that the FLM-F performs almost as well as ANN models for long time forecasting. Figure \ref{fig.FLMFCVcurvesMean} provides an explanation for this by showing the distribution of the CV along the time domain, for the mean over all years. The CV of ANN models is shown to be increasing with the horizon, meaning that the accuracy of the forecasting decreases with the forecasting horizon. This is due to the fact that it is not able to predict whole events. Conversely, the CV does not increase for the FLM-F and is even smaller during the fall season. Figure \ref{fig.FLMFCVcurvesMean} also shows the main drawback of the FLM-F which is that it does not predict well the short-term variations. Indeed, during July streamflows are often low and thus are very dependant on day-to-day rainfall. This is when the mean error of the FLM-F is the highest. Summer streamflows are more difficult to forecast than fall streamflows using only precipitations. This is in agreement with the shape of the $\hat{\beta}(.,.)$ surface which has less amplitude for summer streamflows.

\section{Conclusions}

The purpose of the present work is to introduce and adapt functional linear models to the hydrological framework. This work follows the paper of \cite{Chebana12} which shows the relevancy of using the functional framework in hydrology. After an introduction to FLM models, they are applied to streamflow forecasting based on precipitation curves. Conceptually, FLM are perfectly suited for time series regression since they provide a solution to the problems caused by autocorrelation and non stationarity in time series. Moreover, section \ref{sec.background} shows that elegant estimation methods have been developed to manage infinite dimensional data.

The application of the FLM-S to forecast streamflow volumes provides interesting insights for the interpretation of the results. Indeed, results suggest that precipitations influence streamflow especially in the middle of summer and at the beginning of the fall season. Moreover, this model outperforms the somewhat simple linear regression, in terms of volume forecasting accuracy. The shape of the influence of precipitations on streamflows is refined with the application of the FLM-F. This model highlights the large influence of July precipitations on streamflows two-months later. The importance of summer precipitations on the beginning of the fall season streamflows is also obvious. If ANN models show slightly better performances than the FLM-F, the latter show an ability to forecast the global shape of the hydrograph. Moreover, FLM-F performances do not decrease with the increase of the horizon like ANN models. Moreover, the better performances of the ANN model are due to the fact that we have used the observed precipitation data. However, in practice, precipitation forecasts are not as accurate and the long term forecasts of ANN models would be less accurate than in this application.

The application presented in this paper is relatively simple and deals with a single case study. The present work focuses mainly on the method itself, and illustrates it on a single application. However, a large number of other hydrological applications can be considered. For instance, one could orient the application to other hydrologic events such as spring floods or droughts. In such cases, the curves can be streamflows, snowfalls and temperatures during different seasons. Similar applications can also be performed for different regions representing different climates. Setting a FDA application requires a proper definition of the targeted event and time intervals. FDA can also be used for other purposes such as regional estimation at ungauged sites or estimation of missing data (for instance a continuous part of a hydrograph). A large number of applications need to be investigated in the field of hydrology. 

There are a few limitations to the application of FLMs. One of the particularities of the hydrological framework is the importance of peak streamflow values. For functional data, reaching the peaks necessitates an important amount of basis functions. Similarly to the multiple regression framework, FLMs are based on the modelling of mean curves which do not always reach the desired peaks. Doing so necessitates models that use complex curves which could decrease the performances of the model. Furthermore, the complexity allowed by FLMs also depends on the number of available curves. When using long curves such as in the present work, fitting complex models requires a large number of data years which are not always available. 

The discussion presented above leads to a number of perspectives. First, it can be of interest to apply the historical FLM either from \cite{Malfait03} or from \cite{Kim11} and compare the results with the FLM-F. Second, Following the recent development of the \texttt{R} package \texttt{refund}, a wide body of literature on FLMs is emerging. Notably, we can cite a new estimation method based on mixed models which allows the use of several functional covariates in the FLM-F \citep{Ivanescu13}. This method expresses the FLM-F as an additive model in order to be fit efficiently \citep[such as in][for instance]{WoodBook}. Such an estimation method also has the advantage of providing well justified confidence intervals \citep{Goldsmith11}. Also part of \texttt{refund} is the recent development of functional generalized additive models \citep{McLean14}. Third, an important feature of functions as mathematical objects is the possibility to derive them. This can lead to insights on the variation of streamflow processes and can also be a path for the study of curve peaks. A fourth perspective lies in the use of functional autoregressive models \citep[\emph{e.g.}][]{Damon02} in order to forecast future streamflow phenomena using past streamflow curves. Finally, it appears important in the future to take advantage of the emerging body of literature on functional geostatistics \citep[\emph{e.g.}][]{Delicado10,Caballero13,Ignaccolo13} to model the spatial dependence between hydrological sites.

\bigskip
\emph{Acknowledgements} \\
Financial support for this study was graciously provided by the Natural Sciences and Engineering Research Council (NSERC) of Canada and by the Nord-Pas de Calais Regional Council, France. Moreover, the authors would like to thank the international relations ministry of Quebec (ministère des Relations internationales, de la Francophonie et du Commerce extérieur du Québec) and the french general consulate in Quebec (consulat général de France au Québec) for their financial contribution to this France-Québec cooperation project. Finally, the authors are grateful to the authors of the R package \emph{fda} \citep{FDApackage} for the very handy tools they have provided.

\section*{References}
\bibliography{biblio}

\newpage

\FloatBarrier 

\begin{table}
\centering
\begin{tabular}{l | c c l}
Name & Response $Y$ & Covariates $X^{(j)}$ & Reference  \\ \hline
Classical regression & Scalar & Scalar & \citep{Vogel99} \\
FANOVA & Functional & Scalar & \citep{Brumback98} \\
FLM for scalar response & Scalar & Functional & \citep{Stewart13} \\
Concurrent & Functional & Functional & \citep{Hastie93b} \\
Fully functional & Functional & Functional & \citep[chapter 16]{RamsayBook} \\ \hline
\end{tabular}
\caption{Types of functional linear models.}
\label{tab.flm}
\end{table}

\begin{table}
\centering
\begin{tabular}{l | l l}
Criterion & FLM-S & LM \\ \hline
Bias & 0.001 & 0.000 \\
RMSE & 0.808 & 2.327 \\
LOOCV & 0.183 & 0.192 \\
R$^2$ & 0.940 & 0.506 \\\hline
\end{tabular}
\caption{Values of several criteria in order to compare the FLM-S to the LM. For the bias, RMSE and LOOCV, the lower the criterion, the better the model is, and inversely for the R$^2$.}
\label{tab.criteria}
\end{table}

\begin{table}
\centering
\begin{tabular}{l | l l}
Criterion & FLM-F & ANN \\ \hline
Bias & 0.000 & 0.000 \\
RMSE & 0.671 & 0.211 \\
LOOCV & 0.795 & 0.610 \\
R$^2$ & 0.452 & 1.000 \\\hline
\end{tabular}
\caption{Values of several criteria in order to compare the FLM-F to the ANN. For the bias, RMSE and LOOCV, the lower the criterion, the better the model is, and inversely for the R$^2$.}
\label{tab.criteria2}
\end{table}

\newpage
\FloatBarrier

\end{document}